# Inverted-Mode Scanning Tunneling Microscopy for Atomically Precise Fabrication

*Eduardo Barrera, Bheeshmon Thanabalasingam, Rafik Addou, Damian Allis, Aly Asani, Jeremy Barton, Tomass Bernots, Brandon Blue, Adam Bottomley, Doreen Cheng, Byoung Choi, Megan Cowie, Chris Deimert, Michael Drew, Mathieu Durand, Tyler Enright, Robert A. Freitas Jr., Alan Godfrey, Ryan Groome, Si Yue Guo, Sheldon Haird, Aru Hill, Taleana Huff, Christian Imperiale, Alex Inayeh, Jerry Jeyachandra, Mark Jobes, Matthew Kennedy, Robert J. Kirby, Mykhaylo Krykunov, Sam Lilak, Hadiya Ma, Adam Maahs, Cameron J. Mackie, Oliver MacLean, Michael Marshall, Terry McCallum, Ralph C. Merkle, Mathieu Morin, Jonathan Myall, Alexei Ofitserov, Sheena Ou, Ryan Plumadore, Adam Powell, Max Prokopenko, Henry Rodriguez, Sam Rohe, Luis Sandoval, Marc Savoie, Khalil Sayed-Akhmad, Ben Scheffel, Tait Takatani, D. Alexander Therien, Finley Van Barr, Dusan Vobornik, Janice Wong, Reid Wotton, Ryan Yamachika, Cristina Yu, Marco Taucer*

Affiliations: CBN Nano Technologies Inc., Ottawa, Ontario K1Z 1A1, Canada

## Abstract

Scanning Tunneling Microscopy (STM) enables fabrication of atomically precise structures with unique properties and growing technological potential. However, reproducible manipulation of covalently bonded atoms requires control over the atomic configuration of both sample and probe – a longstanding challenge in STM. Here, we introduce inverted-mode STM, an approach that enables mechanically controlled chemical reactions for atomically precise fabrication. Tailored molecules on a Si(100) surface image the probe apex, and the usual challenge of understanding the probe structure is effectively solved. The molecules can also react with the probe, with the two sides of the tunnel junction acting as reagents positioned with sub-angstrom precision. This allows abstraction or donation of atoms from or to the probe apex. We demonstrate this by using a novel alkynyl-terminated molecule to reproducibly abstract hydrogen atoms from the probe. The approach is expected to extend to other elements and moieties, opening a new avenue for scalable atomically precise fabrication.

# Introduction

Ever since the first demonstration of atomic positioning of xenon atoms on nickel[1], the Scanning Tunneling Microscope (STM) has been used for atomically precise fabrication. The first demonstrations used physisorbed atoms and molecules, but it was soon realized that covalent bonds could also be manipulated. An early example is the selective removal of hydrogen atoms from a passivated silicon surface using inelastic tunneling electrons[2], which remains an important approach for making atomic scale devices[3,4]. After hydrogen removal, samples can be exposed to precursor molecules that react to the surface selectively at depassivated sites. This has been used to embed dopants with nearly atomic precision, making single-atom single-electron transistors[5,6], quantum gates[7,8], and quantum simulators[9,10]. Truly atomic precision in this approach, however, contends with the random orientation of precursor molecules when they impinge on the surface from the gas phase. Several approaches aim to overcome this challenge[11], but so far atomically precise placement has only been shown for a small number of precursor molecules and elements, including phosphorus and arsenic[4,12].

A different approach to atomically precise fabrication can be found in on-surface synthesis. Organic molecules are constrained to a metallic surface and can undergo reactions induced by light, heat, or an STM tip[13–15]. Carbon-halogen bonds can be broken by inelastic tunneling electrons, creating reactive organic species that can be manipulated laterally until, in close enough proximity, they undergo chemical reaction[16,17]. Unlike in the case of adsorption from the gas phase, here the reagents are directed using the STM's lateral degrees of freedom. However, precisely controlling reagent positions using the STM remains difficult in many cases of interest[18]. This approach is also limited primarily to metallic surfaces, which poses a challenge for creating optical or electronic devices, and it is difficult to apply to non-planar atomic structures.

Placing the reagents of a chemical reaction on either side of the tunnel junction allows their relative positions to be controlled in all three spatial dimensions, which, for example, is used for vertical manipulation[19,20]. This overcomes the limitations of two-dimensional on-surface synthesis and also overcomes the uncertainty in reagent orientation that is inherent to gas-phase depositions. But achieving the requisite control of the configuration of atoms on both sides of the tunnel junction is a longstanding challenge of STM. Many probe preparation methods involve contacts with the surface, introducing randomness in composition and atomic coordination. Functionalizing such probes with moieties picked up from the surface can provide some control and even a means of verifying the probe's atomic configuration[21–24], but does not enable reproducible covalent chemical reactions at zero bias.

In this article, we present a new technique, inverted-mode STM, which provides control of both sides of the tunnel junction, enabling reproducible chemical reactions at the tunnel junction. Tall, rigid, and sharp molecules on the sample surface image an atomically flat and crystalline silicon probe apex. The same molecules can also act as reagents. As an example, we demonstrate abstraction of a single hydrogen atom from the H:Si(100)-2x1 surface at zero bias using mechanical control, but our approach generalizes to a wide array of chemical reactions including donation of moieties to the probe apex. Since multiple molecules can sequentially address the same site on the probe apex, this technique will enable atomically precise fabrication of complex, covalently bonded structures.

Figure 1 depicts the experimental conditions that enable hydrogen abstraction in inverted-mode STM. The probe in Figure 1a is terminated in a flat atomic terrace. On the other side of the junction is a novel molecule, EAOGe-C2I, containing an iodoalkyne at its topmost point (see Methods and Supplementary Section 1). Removal of the capping iodine atom leaves a reactive radical, and we refer to such deiodinated molecules as being “activated”. Figure 1b-d shows an atomistic representation of a tunnel junction in which an activated molecule faces a hydrogen-passivated silicon probe apex (the H:Si(100)-2x1 surface). The radical termination of the molecule is laterally aligned with a single hydrogen atom of the probe in Figure 1b. When the molecule approaches the probe surface, it abstracts a hydrogen atom (Figure 1c), and as it retracts the hydrogen atom remains bonded to the molecule (Figure 1d).

The mechanical motion of the probe relative to the sample constrains and steers the reagents as they seek to minimize energy during the chemical reaction. Figure 1e shows the calculated potential energy as a function of the hydrogen atom’s position for two different probe-sample separations. The transfer of the hydrogen atom from the silicon atom of the probe to the carbon atom of the molecule is thermodynamically favored by about 2 eV, in agreement reported bond dissociation energies[25,26] . At separations that are too large (dashed line), an activation barrier prevents the transfer, but as the separation is reduced, this barrier diminishes and eventually vanishes (solid line), guaranteeing hydrogen transfer even at cryogenic temperatures. The reaction proceeds because of 1) the energetics of the reaction, defined by the reagents on either side of the junction, and 2) their relative positions, which are controlled with sub-angstrom precision by the STM scanner; no application of bias or flow of current is required. This ability to achieve specific chemical products by controlling reagents with the STM’s mechanical degrees of freedom can be called mechanosynthesis[27].

Different pairs of reagents (a different molecule and/or a chemically distinct location of the probe apex) will change the energetics of the reaction, which can favor donation of atoms from the molecule to the probe instead of abstraction, or rearrangements without transfer. Sequential reactions in the same region of the probe apex, using a sequence of individual molecules, would therefore amount to the ability to perform atomically precise fabrication on the probe. The promise of this capability is clear, but it depends upon overcoming several key challenges: reproducible fabrication of atomically clean probes, experimental verification of the apex’s atomic structure, a means to align molecules to the probe apex with atomic precision, the ability to characterize reaction outcomes, and the ability to address the same region of a probe apex with multiple molecules, sequentially. The rest of this article will demonstrate that inverted-mode STM addresses all these issues, using the specific example of hydrogen abstraction as just one capability it enables.

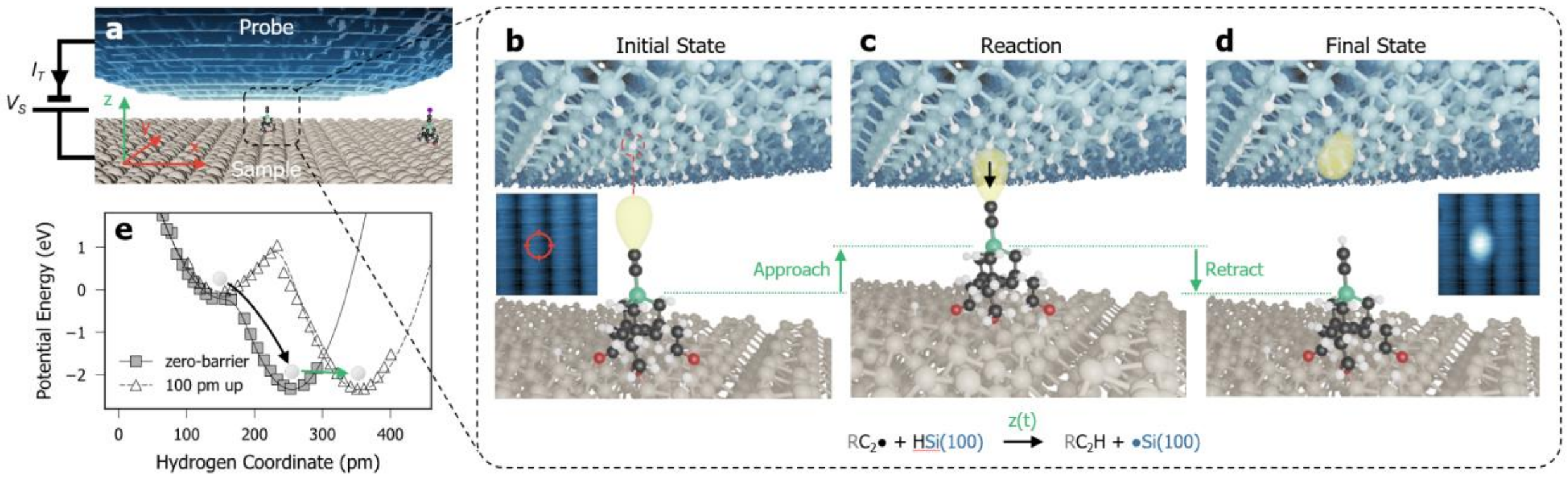

*Figure 1: **Experimental Conditions for Hydrogen Abstraction**. a) Schematic showing the setup. Both probe and sample are Si(100). The probe is hydrogenated, while the sample is unpassivated to enable molecule adsorption. The sample has a sparse coverage of tall molecules (EAOGe-C2I). The x- and y-axes are used to laterally align a molecule with a target location on the build site, while the z-axis can adjust the separation between probe and sample, enabling the reaction to proceed. b-d) Zoom-in on the atomic configuration at the site of the reaction. Initially, the molecule is lined up with an atomic site on the probe apex (b); the molecule then approaches the targeted hydrogen atom until the reaction proceeds (c); finally, the molecule retracts, taking the hydrogen atom with it (d). Insets in (b) and (d) show inverted-mode STM images before and after H-abstraction. e) Potential energy of the system, calculated by density functional theory, as a function of the distance between the hydrogen atom and its host silicon atom, for two different separations of probe and sample (see Methods and Supplementary Section 2). For larger separation (triangles), the barrier to hydrogen transfer is about 1.2 eV. At a critical separation (squares), the reaction becomes barrierless. The black arrow represents the energetically downhill hydrogen transfer from the probe apex to the molecule and the green arrow indicates the retraction.*

# Reproducible Preparation of Atomically Clean Probes

Achieving the conditions that allow reproducible chemical reactions at the tunnel junction begins with fabrication of atomically clean, crystalline probes. UV lithography and potassium hydroxide (KOH) wet etching are utilized to fabricate ~200 μm tall pyramids oriented normal to a degenerately doped n-type Si(100) wafer (Figure 2a-b, and Methods). The wafer is then cleaved into 3 × 3 $mm^2$ Silicon Probe Chips (SPCs). Annealing is performed in a custom-built probe holder (Figure 2f) that passes direct current (DC) through the SPC. The arrangement of electrodes restricts the current flow and localizes Joule heating to ensure the highest temperature occurs at the probe location. As etched, the probe apex has an average radius of curvature (ROC) < 20 nm, which increases to ~1 μm after a 5 minute anneal at 1200 °C (Figure 2c-d).

Atomically resolved characterization using a tungsten STM tip (Supplementary Section 3) shows that, after the anneal, the probe is atomically clean (Figure 2e), with a Si(100)-2x1 terrace at its apex measuring tens of nanometers across. We have found that the annealed probe's ROC increases with annealing, but saturates around 1.9 μm after about 1 hour (Figure 2g), broadly consistent with results from comparable studies[28]. Reduced surface conductivity due to dopant migration during extended anneals is ruled out based on a comparison between short and long anneals showing negligible change in the apparent bandgap at the apex (Figure 2h)[29]. These observations show that the SPC can be annealed repeatedly—at least a dozen times—producing an atomically clean apex of similar dimensions and electronic character each time. Other methods for preparing crystalline probe apexes in the STM head were explored, for instance using probe-sample contact and application of high bias, but they were poorly controlled in comparison (Supplementary Section 6).

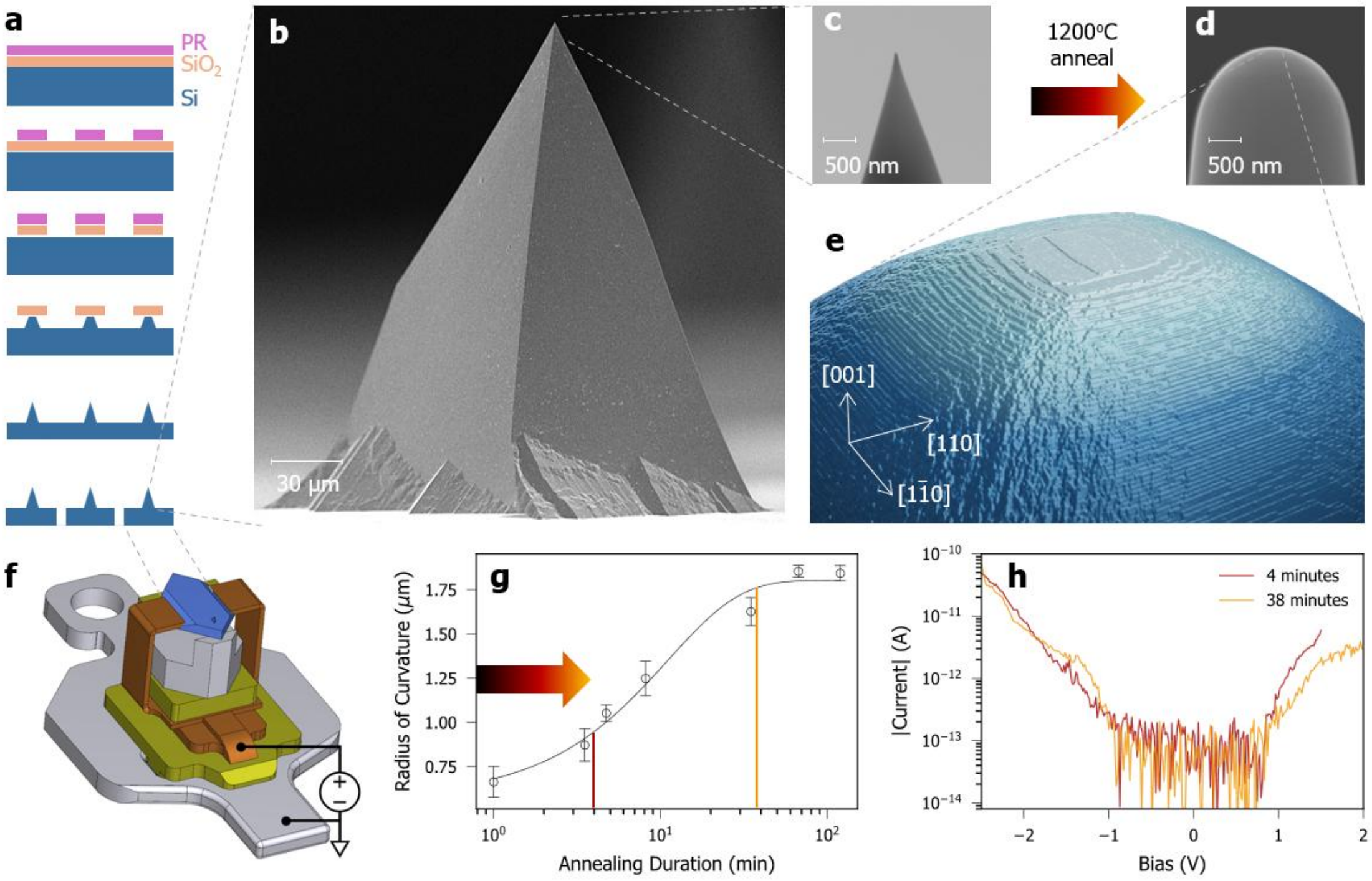

*Figure 2: **Fabrication of an annealable silicon probe chip**. a) Fabrication steps for Silicon Probe Chips (SPCs). A photoresist (PR) is patterned by UV lithography, the pattern is transferred to an $SiO_2$ layer, then pyramidal probes are formed by KOH etching, after which the wafer is cleaved into probe chips. b) SEM image of a silicon probe after fabrication. c-d) Higher magnification SEM images of the apex before (c) and after (d) a 5-minute 1200 °C anneal. e) 3D rendered conventional-mode STM image of an annealed silicon probe, scanned by a sharp tungsten tip, showing atomic terraces with a topmost 66x30 $nm^2$ atomically flat terrace ($V_S$ = -3.5 V, $I_T$ = 30 pA) (Supplementary Section 3). Crystallographic vectors show the [001] direction oriented along z (5 nm length), and the [110] and [1-10] along x and y (50 nm lengths each). f) 3D CAD model of the probe holder with molybdenum clamps that also serve as electrodes holding the SPC in place. g) Radius of curvature of the probe apex vs total annealing time at 1200 °C. h) I(V) spectroscopy of a probe apex after 4 minutes of annealing time and after 38 minutes of annealing time, showing negligible change in the apparent band gap. This measurement was acquired using inverted mode STM where the tunnel junction is composed of a EAOGe-C2I molecule and the Si probe apex (refer to Inverted-Mode Scanning Tunneling Microscopy).*

# Inverted-Mode Scanning Tunneling Microscopy

The molecule used in this work, EAOGe-C2I, is shown as an inset in Figure 3a. It was designed and synthesized for its ability to participate in a range of chemical reactions and for its use in inverted-mode STM. The core of the molecule is a rigid germanium-substituted adamantane cage. Three ethanol groups (colloquially, the "legs") extend from the cage (the "body"). Their terminal OH groups ("feet") can undergo dissociative attachment on the Si(100) surface[30–33], which encourages an upright orientation of the molecule. Bonded to the germanium atom is an iodoalkyne ($C_2I$, the "head"). Details of the molecular deposition procedure can be found in the Methods.

As the annealed probe scans the surface, each protruding molecule provides an image of the SPC probe apex, which we refer to as a Reflected-Probe Image (RPI), several of which can be seen in the large-scale overview image at the bottom of Figure 3a. The coverage of molecules is chosen so that RPIs are generally isolated, but their random distribution occasionally leads to overlapping RPIs. In regions of overlap, multiple molecules tunnel into different parts of the probe apex at the same time, similar to a “double tip” artifact in conventional STM. A subset of fainter RPIs are produced by shorter protrusions on the sample, which may arise from defects or from undesired orientations of the deposited EAOGe-C2I (i.e., “head-down”). Moving from one imaging molecule to another is like swapping tips in conventional STM, but with the added advantage that we can immediately use the new molecule to look at the same atomic region of the probe, which allows comparison of many imagers when characterizing the same site. It is also easy to return to a previous imager simply by moving back to a previously visited molecule.

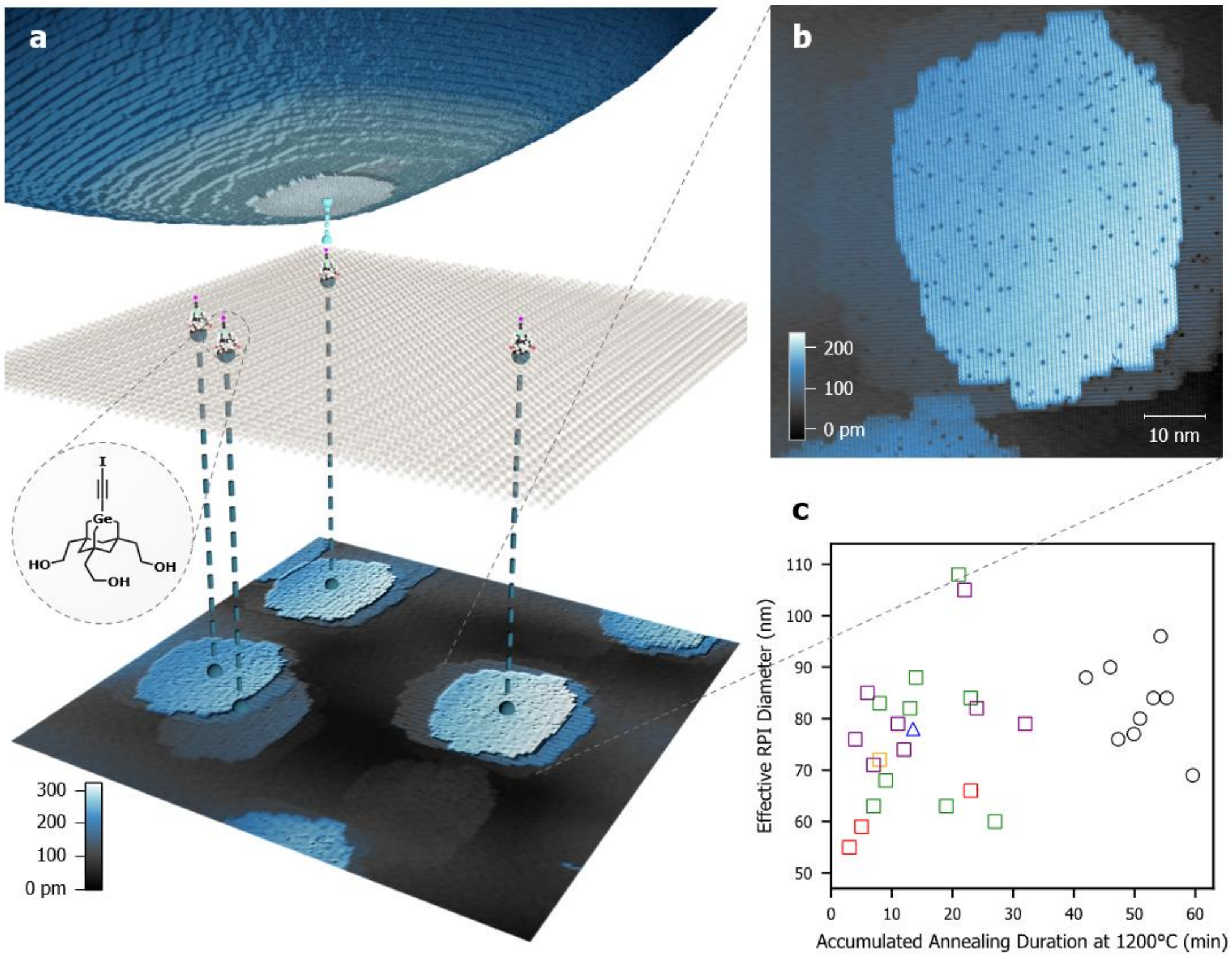

*Figure 3: **Inverted-mode STM**. a) 3D representations of the elements involved in inverted-mode imaging. The probe apex (top) is shown as a 3D rendered illustration based on a composite of multiple real datasets. The sample (middle) is shown schematically, with four molecules serving as imagers. A 200 x 200 nm² overview STM image (bottom) is shown ($V_S$ = -3.5 V, $I_T$ = 30 pA). As the probe scans the sample, each molecule creates an RPI. The inset shows the structure of the EAOGe-C2I molecules used. b) Zoom in on one RPI ($V_S$ = -3.6 V, $I_T$ = 30 pA). c) Effective RPI diameter, $d = \sqrt{4A/\pi}$, where A is the measured RPI area, as a function of accumulated annealing duration at 1200 °C for three different STM systems (represented by the shapes) and for several different SPCs (represented by colors). Most SPCs were annealed several times, producing clean apexes each time.*

Zooming in on a single RPI, as in Figure 3b, we see that the surface-bound molecule images the entire apex terrace of the probe, as well as the terrace below it. This single molecule can access a region of the probe apex up to 100 nm across, including multiple terraces. We can thus experimentally verify the composition and structure of the probe apex, and target reactions at atomically precise sites on it. The ability to see the entire apex terrace in the RPI depends on a relatively small tilt between the crystallographic ⟨100⟩ directions of the probe and sample, but larger tilt angles up to about 2° can be tolerated (Supplementary Section 5). RPIs of annealed probes provide clean step-free regions suitable for atomic fabrication across different STM systems and different total anneal times, and probes can be refreshed multiple times, as shown in Figure 3c.

# Mechanically Controlled Chemical Reaction

To demonstrate a mechanically controlled chemical reaction in inverted-mode STM, we hydrogenate the probe apex (Methods and Supplementary Section 8) and show that hydrogen atoms can be reproducibly abstracted at zero bias. After performing an overview STM scan, we select a small region (a few nanometers across) within an isolated RPI, which we refer to as a build site, shown in Figure 4a. Starting from the initial scanning bias of 3.2 V, we activate the molecule (that is, remove its capping iodine atom) by ramping up the bias with the z-controller feedback on until, at a bias in the range from 3.8 to 4.5 V the height of the RPI increases discontinuously, by around 200 pm. At the same time, imaging characteristics of the molecule change, including a decrease in resolution and qualitative changes like the way step edges look, as seen in Figure 4b. Importantly, the same changes to height and imaging quality are observed after exposing the molecules to UV light. Since carbon-iodine homolytic bond scission is known to be induced by UV light[34,35] , this strongly supports our interpretation of the molecule's orientation and of tip-induced deiodination (Supplementary Section 7).

In Figure 4, the targeted hydrogen atom is four atoms to the right of a naturally occurring dangling bond on the same side of the row. Despite the activated molecule's inability to resolve dimers, it is still possible to align the molecule to the surface with atomic precision. We use the dimer-resolved image of Figure 4a to define surface lattice vectors. Then, to account for any lateral offset that may occur during activation, we use the row-resolution of both terraces visible in Figure 4b to laterally align the lattice grid. The experimental data therefore completely constrains the surface lattice, such that atomic positions can be determined. Sequential images of the build site are used to correct for piezo creep and drift to first order. We use the lateral offset between forward- and backward-scans to correct for the hysteresis in scan direction.

With atomically precise alignment of the molecule's terminal radical to a specified hydrogen, we can perform mechanosynthesis. The probe-sample separation is initially held fixed and the bias is set to zero so that no current flows. Since both the sample and the probe are degenerately doped n-type silicon, the contact potential, and consequently the electric field, is small. The chemical reaction is thus dominated by the reactivity of the ethynyl radical, which strongly favors hydrogen abstraction[36]. The barrier to hydrogen transfer is then controlled by adjusting the probe-sample separation (as in Figure 1).

The molecule approaches the hydrogen atom by a specified distance and then retracts. The build site is then scanned again with the same molecule to determine if a change occurred. The initial approach is 200 pm deep, referenced to the tunneling setpoint, and the procedure is repeated with sequential approaches 50 pm deeper each time. At a 350 pm approach, three changes are observed simultaneously (Figure 4c): first, there is a change in the image height; second, the image's qualitative appearance changes; and third, we see a new protrusion in the inverted-mode STM image of the probe apex at the interaction location. All three observations are consistent with transfer of a hydrogen atom from the probe apex to the molecule, leaving a dangling bond on the probe and a terminal $C_2H$ group on the molecule. This outcome – removal of a single hydrogen atom at zero volts – was observed in 27 out of 28 trials, indicating that the process is well controlled and reproducible (Methods and Supplementary Section 10).

Finally, moving to a different RPI, we can image the same build site with an iodinated molecule (Figure 4d). The image is similar to the first one produced by the target molecule (Figure 4a), consistent with both molecules' initially iodine-capped state, except for the new dangling bond at the targeted location. The after-image is not only useful to characterize the change to the build site, it also shows that the build site can immediately be addressed by another molecule. That is, the same procedure can be repeated with this new molecule to remove another hydrogen atom. Moreover, by changing either the molecule or the build site, sequential reactions of different types become possible. Because of the EAOGe-C2I molecule's relatively weak Ge-C bond[37], other chemical reactions that can be expected for a suitable state of the build site (*e.g.*, different arrangements of dangling bonds) are donation of $C_2$, $C_2I$, or $C_2H$ groups to the probe apex, with all three of these groups demonstrated as imagers in Figure 4. Other molecules could be designed to transfer different moieties in either direction. We thus see that inverted-mode STM enables a far-reaching capability for multi-step atomically precise fabrication by sequential chemical reactions.

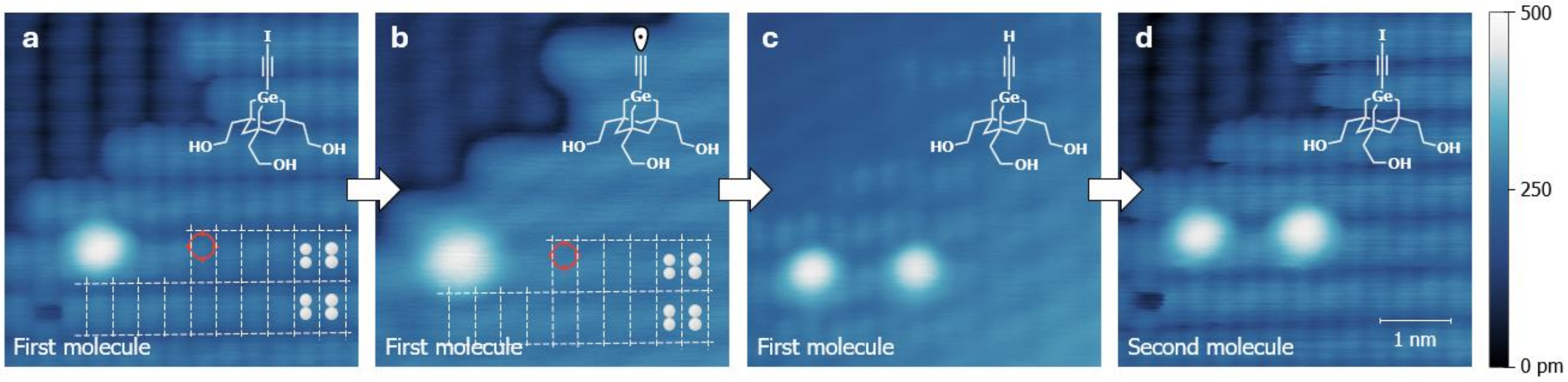

*Figure 4: **Mechanosynthetic hydrogen abstraction**. a) STM image of the selected build site before H-abstraction, imaged by a EAOGe-C2I molecule, capped by an iodine atom ($V_S$ = 3.2 V, $I_T$ = 10 pA). Surface unit cells are outlined by dashed lines, locations of hydrogen atoms are schematically illustrated in the bottom right, and the targeted location is indicated by a red bull's eye. b) STM image of the same region after deiodination of the molecule ($V_S$ = 3.8 V, $I_T$ = 10 pA). c) STM image of the build site after the reaction, using the spent molecule, now capped with a single hydrogen atom ($V_S$ = 3.2 V, $I_T$ = 10pA). d) STM image of the same build site imaged by a different molecule that is capped by iodine, showing the same imaging characteristics as the first molecule when it was in the same state ($V_S$ = 3.2 V, $I_T$ = 10 pA). Inset molecular structures indicate the state of the molecule used to image the build site. The topographic color scale spans 500 pm in all cases.*

# Conclusions

We have shown that inverted-mode STM provides a new way to control and verify the atomic structure on both sides of a tunnel junction. Individual molecules, sparsely deposited on the sample, each provide an image of the atomically clean probe apex and can be aligned to it with atomic precision. The probe and sample can be brought together to induce a chemical reaction. The probe can then be addressed by another molecule on the sample, providing means to characterize the outcome of the reaction, and at the same time creating the conditions for a subsequent reaction at the same site.

We expect that similar probe preparation can be expanded to other materials, opening avenues for atomically precise fabrication in new contexts. Applications to other semiconductors should be straightforward. We have seen that metal probes can be conditioned for inverted-mode STM by multiple means in the STM head (Supplementary Section 6). Two-dimensional materials could be used by transferring nanoscale flakes to clean probe apexes by *in-situ* exfoliation, which may enable fabrication of optically active point defects[38].

For characterization applications, the use of surface-bound molecules as imagers can help standardize experimental conditions. Molecules can be selected for their height, conductivity, electronic structure (*e.g.* local density of states), or a specific imaging orbital (*e.g.* s- or p-wave). Since a given probe apex can be addressed by many molecules, comparison of images formed by different types of molecules is possible. Given this diversity, understanding the adsorption configurations of three-dimensional molecules, while tractable, emerges as a key frontier in surface science.

Mechanical control of individual reactions introduces a new lever to control reaction outcomes. This article has presented a simple motion along the *z*-axis to enable H-abstraction, but three-dimensional trajectories can steer the reagents through a higher-dimensional configuration space (Supplementary Section 9). Precisely understanding and controlling the "true" (that is, internuclear) separation between probe and sample emerges as another important challenge. Experimental signals such as tunneling current if bias is applied, or frequency shift in atomic force microscopy, can shed light on these questions, enabling fundamental studies in chemistry.

Atomically precise fabrication represents the most significant application of inverted-mode STM. With exquisite control of both sides of the tunnel junction, a wide set of new atomic manipulations will become possible on silicon, among other material platforms. While we have demonstrated hydrogen abstraction, other subtractive operations are possible, and addition of atoms and molecular fragments to a build site is a topic of ongoing research. This method is applicable to covalent chemistry involving many elements including carbon, and we anticipate its use in constructing increasingly complex atomic structures.

# Methods

## Silicon Probe Chip Fabrication

The fabrication process for Si pyramidal probes is outlined in Fig 2a. The structures were fabricated on a degenerately As-doped Si(100) wafer (4 mΩ-cm) with a thermal oxide on both sides. A ~1 µm-thick photoresist (Shipley S1811) was spin-coated onto 2 × 2 $cm^2$ samples, patterned by UV lithography, and developed in MF-24A. The pattern was transferred into the oxide using a 5:1 buffered-oxide etchant (BOE), followed by resist removal in acetone. The sample was wet etched in 30 wt% KOH at 80 °C to form pyramidal structures. Based on the measured etch rates of Si (1.14 ± 0.02 µm/min) and thermal oxide (7.0 ± 0.5 nm/min) in the KOH solution, an oxide thickness of 1.5 µm was selected as the KOH etch mask to comfortably achieve the target pyramid height of ~200 µm. The KOH etching setup comprised a PTFE beaker placed in a heated oil bath, equipped with a vertical sample holder and magnetic stirring for uniform etchant mixing. A PTFE lid was used to maintain a stable temperature and KOH concentration by minimizing heat and water losses to the surroundings, while allowing temperature monitoring through a tightly fitted opening. Anisotropic KOH etching and corner undercutting enable formation of pyramidal structures on Si(100)[39]. Particularly, in the case of square masks aligned along the ⟨110⟩ direction, Si(111) planes form at the edges of the square, while undercutting at the convex corners of the square mask results in lateral etching and gradual narrowing of the mesa plateau into a pyramid with a sharp apex[40]. The final pyramid morphology can be modified by using specific additives[41–43] or selecting the appropriate etchant type and concentration[44]. Here, we employ a triangular mask aligned to three specific high-etch-rate planes, producing three-sided pyramids[45–47]. These structures self-sharpen to an average apex radius < 20 nm even after the oxide mask falls off and the apex is subject to over-etching[45]. Finally, residual oxide was stripped in BOE, and the wafer was cleaved into individual 3 x 3$mm^2$ SPCs.

Prior work using millimeter-scale substrates as STM or atomic force microscopy (AFM) probes – including devices with annealing capability[48] – have been explored[49–53]. However, annealing at temperatures near 1200 °C was generally not pursued in those cases, nor the ability to obtain a crystalline surface at the probe apex. Our approach relies on high temperature annealing as a key step that removes native oxide and carbon contaminants, yielding an atomically clean silicon build site for mechanosynthesis. During the anneal, diffusion of Si adatoms from terrace step edges leads to a "step flow" that promotes the formation of a wide step-free topmost terrace[54] – a process associated with the mesoscopic blunting of the probe[28,55]. Annealing is performed in a custom-built probe holder (Fig. 2f) comprised of two molybdenum (Mo) clamps that secure the SPC against a Mo block. Ceramic spacers provide electrical isolation and enable direct current (DC) to pass from the two clamps on the SPC into the grounded block through the front corner of the SPC. This arrangement restricts the current flow and localizes Joule heating, ensuring the highest temperature occurs at the probe location and enabling efficient DC annealing. Alternative in situ probe conditioning has demonstrated successful creation of crystalline apex structures, as evidenced in prior reports[56–58] and in our early work with Si and W probes (see *Explorations of Using Inverted Mode STM for In-situ Probe Preparation*). However, this approach does not yield a reliable and reproducible recipe. By contrast, DC annealing provides a robust pathway to atomically clean apices and should extend to other semiconductors and metals after an optimization of annealing parameters.

## Sample Preparation and STM Measurements

The silicon samples are made from degenerately arsenic-doped Si(100) wafers, with room-temperature resistivity in the range 1-5 mΩ-cm. Samples were loaded into a UHV system with base pressure below $3 \times 10^{-10}$ mbar. They were outgassed at 650 °C for several hours, then annealed at 1200 °C for 5 minutes or more by resistive heating using a DC electrical current. Before molecular deposition, the sample, sample holder, and manipulator are allowed to cool for at least 20 minutes.

Molecules were deposited using an effusion cell in a separate chamber from the one where samples are annealed. A quartz crucible, loaded with EAOGe-C2I, was heated to 39°C and held at that temperature for at least 30 minutes to ensure a stable temperature and to outgas the effusion cell. Just before the deposition, the sample was flash-annealed to 1200°C for 10 seconds and allowed to cool for 60 seconds. The sample was then brought into the deposition chamber. A shutter on the effusion cell was then opened to start the molecular deposition, which lasted 5 minutes. After the deposition, the sample was immediately transferred into the STM head. This procedure was used for the sample in Figure 4, but the specific deposition temperature and time was adjusted slightly in other experiments to ensure a suitable coverage of molecules.

Most results presented used annealed silicon probes. When a sharp metallic probe was needed for preliminary testing in “conventional-mode” STM (as in the case of Figure 2e), tungsten probes were used. They were prepared by sharpening a polycrystalline W(110) wire via electrochemical etching in a ~5 M NaOH solution, followed by conventional in-situ tip-conditioning procedures to reach atomic level sharpness.

The statistics on success rate for hydrogen abstraction reported in *Mechanically Controlled Chemical Reaction* and expanded upon in the Supplemental section *Additional Data on Hydrogen Abstraction on H:SPC* were collected with a procedure similar to the one reported in Figure 4. For each mechanosynthetic interaction, the molecule approached the probe by a set distance, and if no change was observed, the approach was repeated at increasing depths until a change was observed ensuring the hydrogen transfer occurred at a minimal approach depth. For some trials, however, the initial depth of approach led to hydrogen transfer on the first attempt, indicating that the approach was likely deeper than needed.

All STM experiments were performed at liquid-helium temperature (4.4 K) using a bath cryostat and a custom-built STM head.

## Computational Modeling

Density functional theory (DFT) calculations were performed with the B3LYP hybrid density functional[59], 6-311G(d,p) basis set[60], and D3 version of the Grimme dispersion correction with Becke-Johnson damping[61] in TeraChem 1.96H[62,63] with a wavefunction convergence threshold of 1.0e-8 a.u and two-electron integral threshold of 1.0e-12. A mesh of energy sampling points for the H-atom positions was generated by constraining select atoms in a reduced molecular model and a small Si(100) proxy to X,Y,Z positions as shown in Supplemental Figure S1.

## Acknowledgements

The work presented in this manuscript was funded by the Canadian government under the Strategic Innovation Fund (SIF, Agreement Number 813022), and by Canadian Bank Note Company, Ltd. (CBN) under "CBN Nano Technologies, Inc." We thank Matthew Koko-Smith and Garrett Thompson for providing technical and hardware support. We wish to thank Amnaya Awasthi and Vaibhav Thakore for aid in literature search. We thank Bill Cullen for contributions to hardware design and development of the SPM head used in this work. We thank John Su and Sam Shields for chemistry support. We thank Sarah Burnside for technical support related to software. We thank Darian Blue for organizational support in preparation of this publication. We thank Steven DeSmet for project and program management. Access to the NanoFab facilities at the University of Ottawa and their staff is greatly appreciated.

## Author Contributions

AI, BB, BT, DV, EB, JB, KSA, MD (Drew), MD (Durand), MK (Kennedy), MT, RAF, RCM, RY, SL, and TT contributed to conception of the work. AI, BB, BC, BT, EB, KSA, MC, MD (Drew), OM, and RG contributed to experimental design. AA, AB, AG, AH, AI, AM, AO, BB, BC, BS, BT, CD, CI, CJM, DA, DAT, DC, EB, FVB, HM, JB, JM, JW, KSA, LS, MC, MD (Drew), MK (Kennedy), MK (Krykunov), MM, MS, MSM, MT, OM, RA, RAF, RCM, RG, RJK, RP, RW, SH, SL, SR, TE, TH, TM, and TT contributed to experiment, theory, and hardware research and development. AA, AH, AI, AP, BT, CD, CJM, DAT, EB, FVB, HM, JM, JW, KSA, MD (Drew), OM, RG, SL, SYG, TE, and TH contributed to data analysis and interpretation. AI, AP, CJM, DAT, FVB, JJ, JM, MJ, MP, RG, SO, and TB contributed to software development. EB, BT, and MT wrote the manuscript with contributions and revisions from AB, AI, BB, BC, CJM, DA, HM, JB, KSA, MC, MJ, OM, RJK, RP, RW, RY, SR, SYG, TE, and TH.

# Supplementary Information: Inverted-Mode Scanning Tunneling Microscopy for Atomically Precise Fabrication

*Eduardo Barrera, Bheeshmon Thanabalasingam, Rafik Addou, Damian Allis, Aly Asani, Jeremy Barton, Tomass Bernots, Brandon Blue, Adam Bottomley, Doreen Cheng, Byoung Choi, Megan Cowie, Chris Deimert, Michael Drew, Mathieu Durand, Tyler Enright, Robert A. Freitas Jr., Alan Godfrey, Ryan Groome, Si Yue Guo, Sheldon Haird, Aru Hill, Taleana Huff, Christian Imperiale, Alex Inayeh, Jerry Jeyachandra, Mark Jobes, Matthew Kennedy, Robert J. Kirby, Mykhaylo Krykunov, Sam Lilak, Hadiya Ma, Adam Maahs, Cameron J. Mackie, Oliver MacLean, Michael Marshall, Terry McCallum, Ralph C. Merkle, Mathieu Morin, Jonathan Myall, Alexei Ofitserov, Sheena Ou, Ryan Plumadore, Adam Powell, Max Prokopenko, Henry Rodriguez, Sam Rohe, Luis Sandoval, Marc Savoie, Khalil Sayed-Akhmad, Ben Scheffel, Tait Takatani, D. Alexander Therien, Finley Van Barr, Dusan Vobornik, Janice Wong, Reid Wotton, Ryan Yamachika, Cristina Yu, Marco Taucer*

Affiliations: CBN Nano Technologies Inc., Ottawa, Ontario K1Z 1A1, Canada

# 1. Synthesis and Characterization

All reactions were performed in Pyrex glassware or vials, equipped with a stir bar, and capped with a septum. For reactions using inert atmospheres, oven-dried glassware was cycled under vacuum and argon atmosphere (3–5 times). Commercial reagents were used without further purification. Tetrahydrofuran was dried and stored over molecular sieves (4Å, 8–12 mesh). Unless otherwise stated, yields refer to products that were isolated after purification. Reaction progress was tracked using thin layer chromatography (TLC) analysis and visualized with UV light (254 nm) and/or with stains ($KMnO_4$, PA, $I_2/SiO_2$). Automated flash column chromatography was performed using a Yamazen Smart Flash equipped with an ELSD and UV detector (254 nm). $^1H$ NMR (600 MHz) and $^{13}C$ NMR (151 MHz) were recorded on a Bruker Avance III HD 600 spectrometer at the University of Ottawa (an artifact in $^{13}C$ NMR spectra related to the hardware of the instrument was present at 104.86 ppm in $CD_3OD$ before referencing the solvent; 106.22 ppm after referencing in the spectra below and not identified with peak picking). NMR samples were dissolved in chloroform-d ($CDCl_3$), methanol-$d_4$ ($CD_3OD$), dichloromethane-$d_2$ ($CD_2Cl_2$), or dimethyl sulfoxide-$d_6$ (DMSO-$d_6$; $(CD_3)_2SO$). NMR spectra contained herein were processed using the automated phase correction and baseline correction features available in MNova, followed by referencing the spectra to the appropriate solvent residual. Chemical shifts are reported in ppm, and referenced to the solvent residual ($CDCl_3$: $^1H$ NMR: $\delta$ = 7.26 ppm, $^{13}C$ NMR: $\delta$ = 77.16 ppm; $CD_3OD$: $^1H$ NMR: $\delta$ = 3.31 ppm, $^{13}C$ NMR: $\delta$ = 49.00 ppm; $CD_2Cl_2$: $^1H$ NMR: $\delta$ = 5.32 ppm, $^{13}C$ NMR: $\delta$ = 53.84 ppm; $(CD_3)_2SO$: $^1H$ NMR: $\delta$ = 2.50 ppm, $^{13}C$ NMR: $\delta$ = 39.52 ppm). Peak multiplicities were described accordingly: s = singlet, d = doublet, t = triplet. Infrared (IR) spectra were recorded on an Agilent Cary 630 FT-IR spectrometer with an ATR sampling module. Spectra were obtained from neat samples (thin films or solids), with characteristic absorption wavenumbers ($\nu_{max}$) reported in $cm^{-1}$. Peaks pertaining to infrared spectroscopy are described accordingly: w = weak, m = medium, s = strong, vs = very strong, br = broad. High-resolution mass spectrometry (HR-MS) analysis was performed by the John L. Holmes Mass Spectrometry Facility at the University of Ottawa on a Kratos Concept 2S (electron impact (EI) ionization, magnetic sector analyzer).

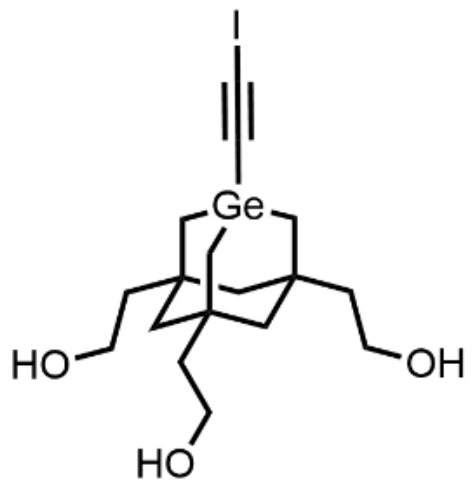

**2,2',2''-(1-(iodoethynyl)-1-germaadamantane-3,5,7-triyl)tris(ethan-1-ol): EAOGe-C2I**

The title compound was synthesized according to a recent patent [27]

**IR** (neat, $cm^{-1}$): 3314 (br), 2900 (m), 2845 (m), 2090 (m), 1645 (w), 1405 (m), 1345 (w), 1219 (w), 1102 (w), 1067 (m), 1014 (s), 807 (m), 693 (s).

**HR-MS** (EI): *m/z* calc'd for $C_{15}H_{22}GeIO_2$ $[M–CH_3O]^+$: 432.9885, found: 432.9863.

**$^1$H NMR** (600 MHz, $CD_3OD$) δ = 3.66 (t, *J* = 7.5 Hz, 6H, 3 X $CH_2$, $CH_2$–$\underline{CH_2}$–OH), 1.46 (t, *J* = 7.7 Hz, 6H, 3 X $CH_2$, C–$\underline{CH_2}$–$CH_2$–OH), 1.39 (d, *J* = 13.1 Hz, 3H, 3 X CH, C–$\underline{CH_2}$–C, ring: equatorial), 1.11 (d, *J* = 13.2 Hz, 3H, 3 X CH, C–$\underline{CH_2}$–C, ring: axial), 0.92 (s, 6H, 3 X $CH_2$, C–$\underline{CH_2}$–Ge) ppm.

**$^{13}$C{$^1$H} NMR** (151 MHz, $CD_3OD$) δ = 96.6 (C, Ge–$\underline{C}$≡C), 58.6 (3 X $CH_2$, $CH_2$–$\underline{CH_2}$–OH), 52.1 (3 X $CH_2$, C–$\underline{CH_2}$–$CH_2$–OH), 50.1 (3 X $CH_2$, C–$\underline{CH_2}$–C, ring), 37.6 (3 X C, $\underline{C}(CH_2)_4$), 25.7 (C, C≡$\underline{C}$–I), 24.0 (3 X $CH_2$, C–$\underline{CH_2}$–Ge) ppm.

**NMR Spectra**

**$^{1}$H NMR** (600 MHz, $CD_3OD$): **EAOGe-C2I**

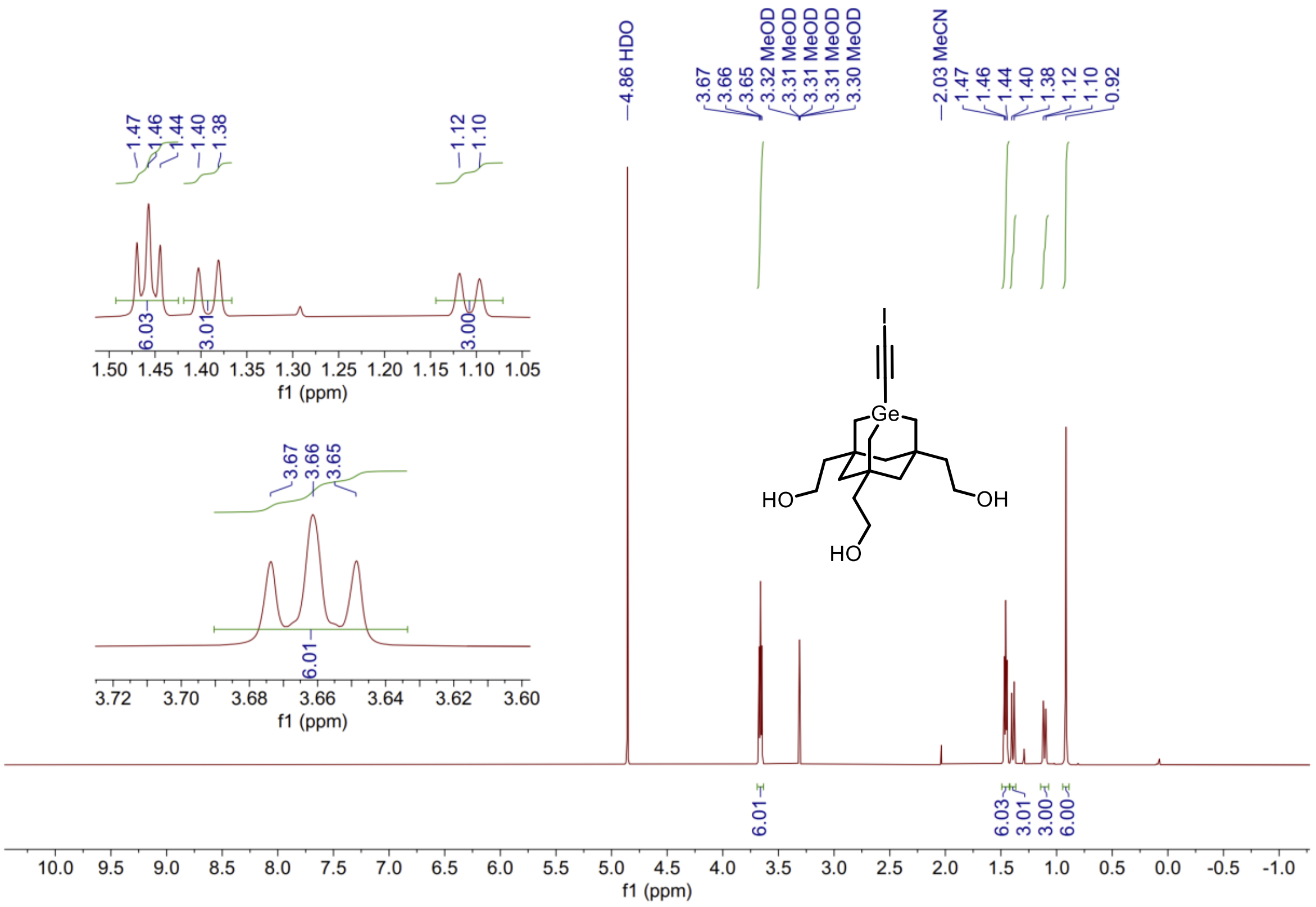

$^{13}$**C{$^{1}$H} NMR** (151 MHz, $CD_3OD$): **EAOGe-C2I**

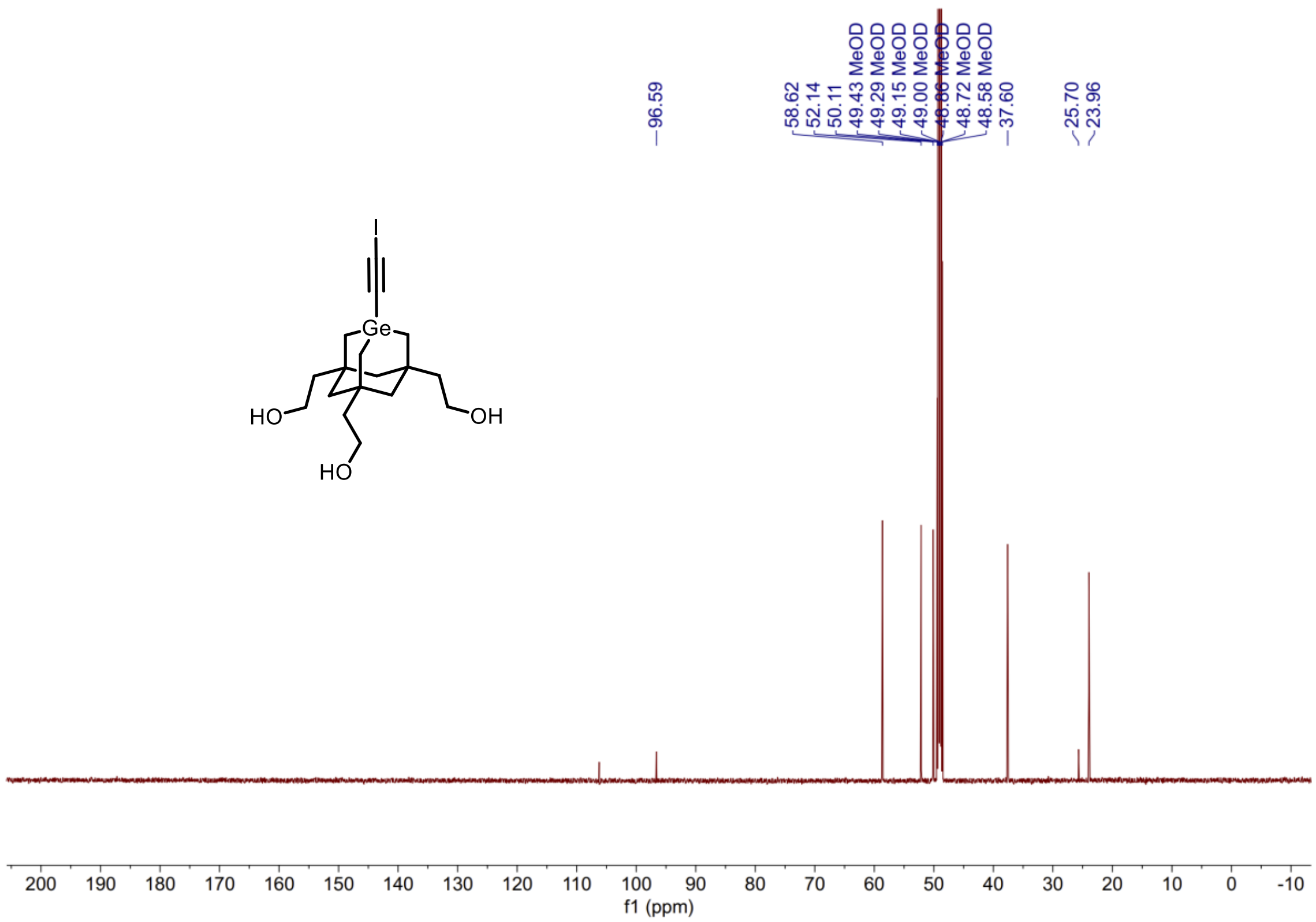

**HSQC** (DEPT-135 modified; $CD_3OD$): **EAOGe-C2I**

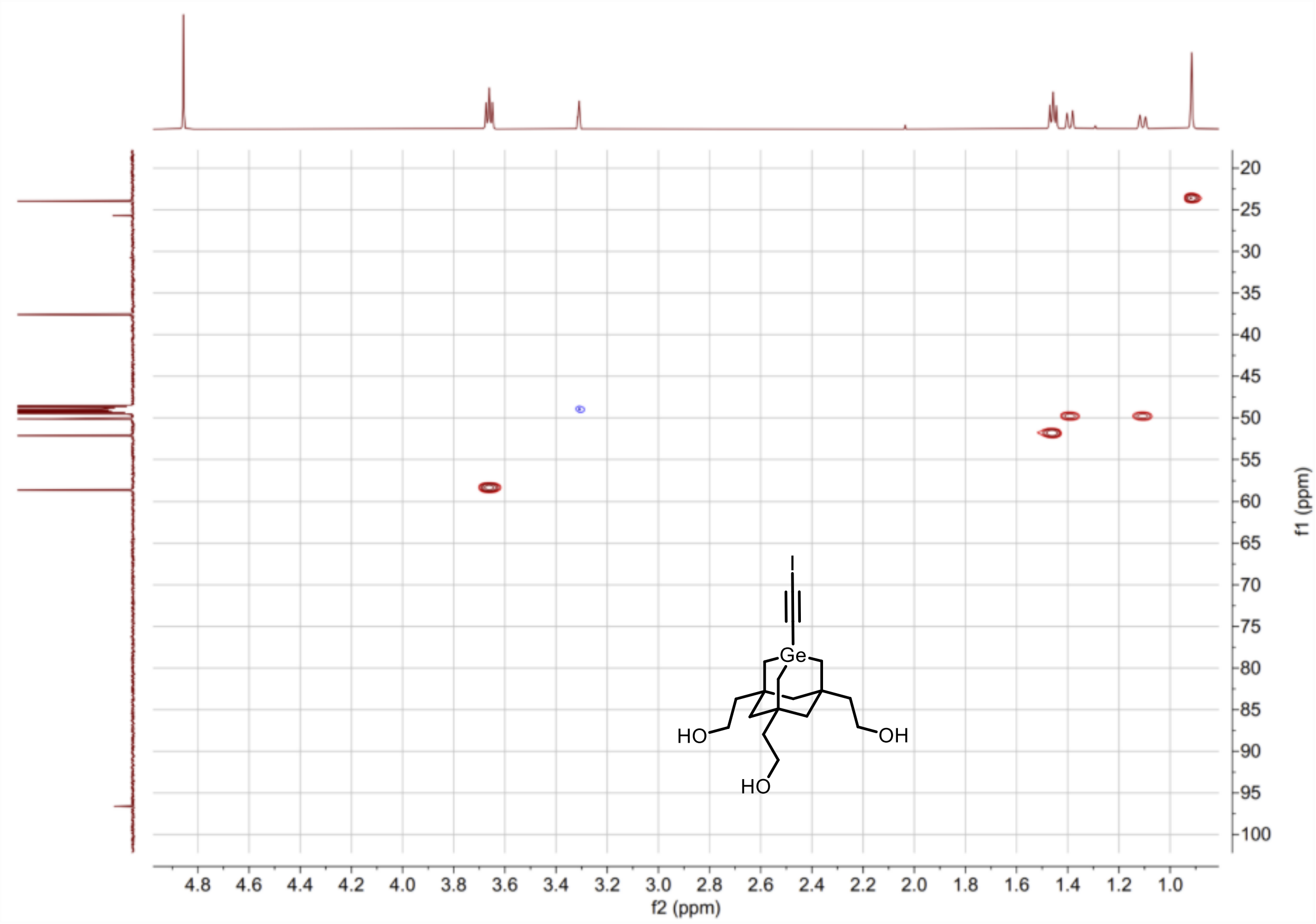

## 2. Computational Modeling

The atomic configurations used for calculations of potential energy are shown in Figure S1. Refer to the Methods section for computational details.

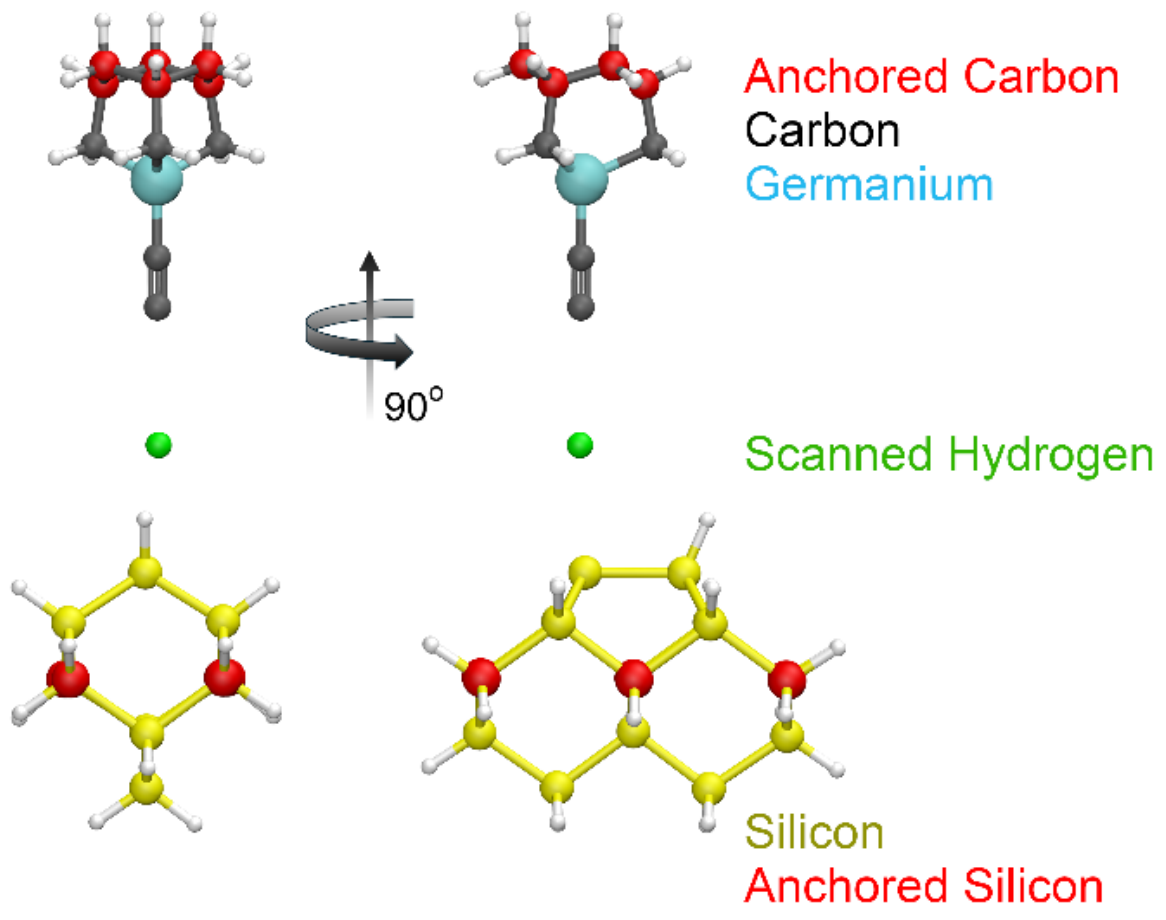

*Figure S1: **Atomic model used for calculations**. The germa-adamantane cage with pendent ethynyl radical and the hydrogen atom scanned (green) to produce the potential energy surface for its abstraction from a model Si100 proxy (anchored atoms in red).*

# 3. Conventional Mode STM Characterization of Annealed Silicon Probes

The effectiveness of direct-current annealing in producing probes with atomically clean apex terraces of suitable dimensions was validated by examining the apex surface morphology of annealed sharp Si protrusions using conventional STM. Comparable STM studies have reported atomically flat, step-free terraces extending from hundreds of nanometers to several micrometers after annealing mesa structures. However, sharp Si probes with ROCs < 20 nm had not previously been investigated under similar annealing conditions. Following the fabrication procedure described in *Silicon Probe Fabrication Methods*, we prepared a Si sample containing a millimeter-scale array of probes approximately 2.5 μm tall—nearly two orders of magnitude shorter than the SPC probes discussed in the main text. A probe pitch of ~4 μm allowed multiple apexes to be imaged within the accessible STM scan area of 10 x 7 $\mu m^2$.

The array was introduced into ultra-high vacuum (UHV), outgassed at 600 °C for ~16 h, and subsequently flash-annealed at 1200 °C for 2 min, yielding a clean Si surface as confirmed by post-STM SEM inspection (Figure S2a). The W probe located the apex of each annealed Si probe by monitoring the local surface slope and advancing in the uphill direction until the topmost terrace was reached. Each examined apex exhibited a clean, atomically flat, and highly conductive Si(100) terrace with a (2×1) surface reconstruction and a lateral extent below ~100 nm.

In contrast, STM imaging of a separate probe array annealed at 1100 °C for 30 s revealed apexes lacking atomic flatness and displaying features tens of nanometers high (Figure S2d–e). These surface irregularities correspond to similarly tall protrusions observed by SEM whose origin is attributed to insufficient annealing. During STM characterization, we also imaged the Si(100) valley regions between adjacent probes (Figure S2f), high-index Si facets such as the one shown in Figure S2g, and the flat apexes of neighbouring probes (Figure S2h-k) – including the 3D STM rendering shown in Figure 1e of the main text. Current-voltage spectroscopy revealed a markedly lower conductivity in the valley regions than at the probe apexes, corresponding to an approximately threefold increase in the apparent Si bandgap (Figure S2l). One possible explanation for this difference is the transfer of silicon from the pyramidal protrusions to the valleys by step-flow during anneal. Since arsenic dopants are known to be depleted from the near-surface region of annealed silicon samples, the silicon that gathers in the valleys could have a lower dopant density compared to the “freshly exposed” silicon at the apexes.

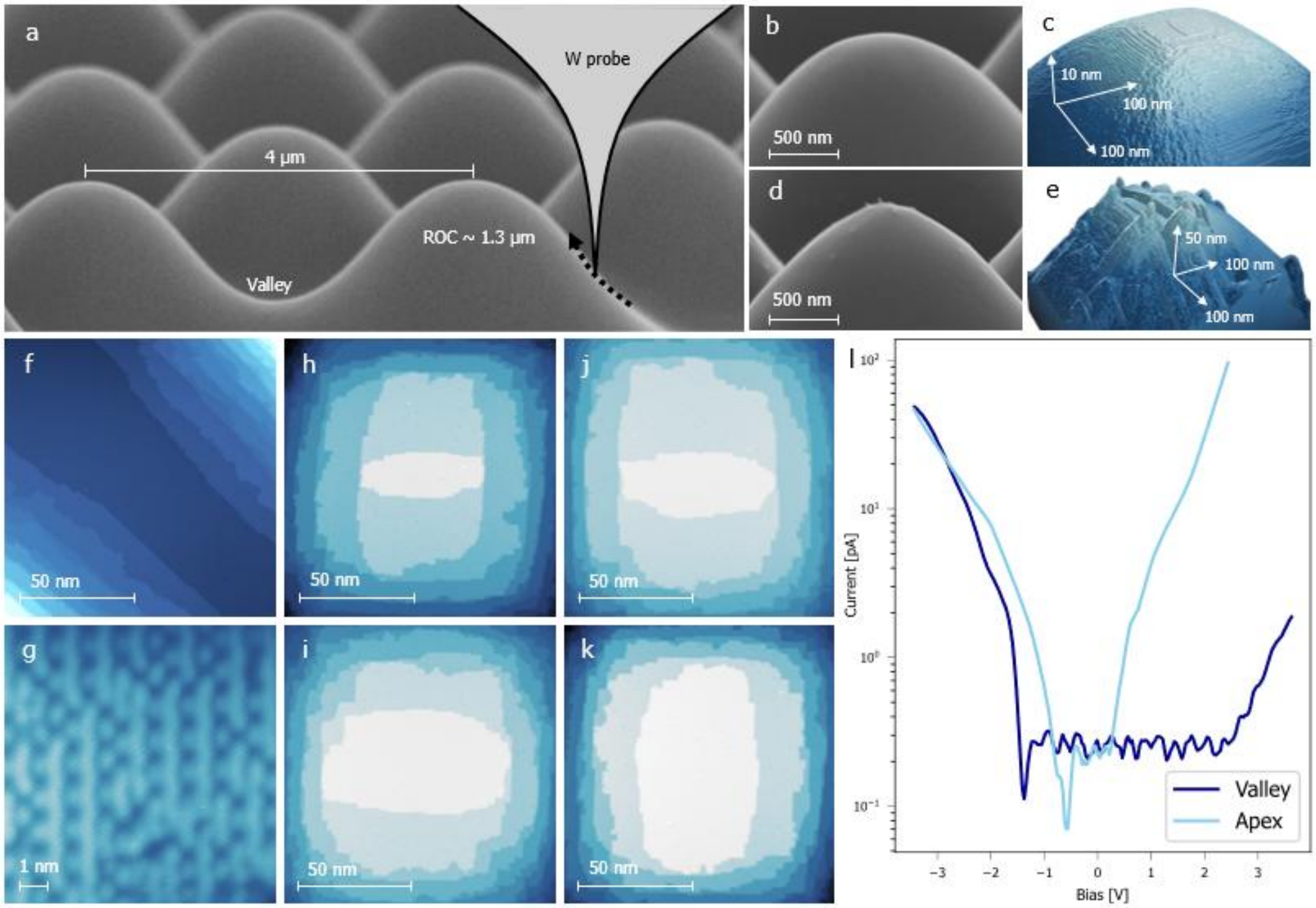

*Figure S2:* ***Atomic Inspection of Annealed Si Probe Apices:*** *a) SEM image of a silicon probe array annealed at 1200 °C for 2 min which is inspected by STM using a sharp W probe. b-c) Higher magnification SEM image (b) and 3D rendered conventional-mode STM image (c) of a probe apex from the array in (a). d-e) Higher magnification SEM image (d) and 3D rendered conventional STM image (e) of a probe apex from an array similar to (a) but annealed at 1100 °C for 30s. Clean and crystalline terraces are present in (c) while a high density of nanometer scale contaminants are observed in (e). f-k) Conventional-mode STM images of the array of Si probes in (a) at various locations: the valley between neighbouring probes (f), a high-index crystal plane on the side of a probe (g), and the top terraces of four distinct probe apices (h-k). l) I(V) spectroscopy using a W probe over the topmost terrace and the valley between two probes, showing a marked increase in the apparent Si bandgap in the latter.*

# 4. Additional Data on Si Probe Annealing

Annealing of the SPCs is conducted under ultra-high vacuum (UHV) using a custom-built receptacle that enables electrical connection between the probe holder and external power supply via finger electrodes. The SPC temperature is monitored by a pyrometer (PYROSPOT DG 10NV, DIAS Infrared Systems), while a rack-mounted (Genesys Series, TDK-Lambda) power supply provides DC current. A programming interface controls the current output to maintain a constant annealing temperature, monitors chamber pressure via a hot-cathode ion gauge, and automatically interrupts and resumes heating to ensure the chamber pressure remains generally between ~$5x10^{-10}$ mbar and ~$2x10^{-9}$ mbar. The program terminates the annealing sequence once the target cumulative duration is reached. Prior to SPC mounting, the probe holder is degassed by resistive heating to reduce residual contaminants, which minimizes water-related defects at the apex after annealing.

Although our standard annealing process involves a 5 min anneal at 1200 °C, lower temperatures (900–1100 °C) offer distinct advantages. Annealing below 1200 °C enables finer control of the apex radius of curvature (ROC) and reduces the build-site area. For example, a 5 min anneal at 900 °C yields an apex ROC that is nearly an order of magnitude smaller than the standard 1200 °C process (see Figure S3b-e). Lowering the annealing temperature also facilitates integration of our mechanosynthesis protocol with industry-standard semiconductor front-end processes, which are typically limited to temperatures below 1200 °C for short durations to avoid excessive dopant diffusion, oxide degradation, and stress accumulation. Likewise, STM lithography and MBE growth face similar thermal-budget limitations when coupled to conventional semiconductor fabrication processes, motivating the development of low-temperature surface-preparation techniques for Si, including ex-situ methods such as protective oxide growth[64], as well as in-situ approaches such as reactive hydrogen treatments[65,66] and sputtering[67], each demonstrating varying degrees of success. Such strategies could be directly implemented in our mechanosynthesis protocol. However, without these low temperature treatments, SPCs annealed below 1200 °C develop persistent nanoscale surface features that prevent the formation of atomically clean and flat build sites. In contrast, a 5 min anneal at 1200 °C yields an atomically clean, defect-free apex (Figure S3e).

The probe aspect ratio (AR) also strongly influences the final apex morphology. Probes with low AR (roughly <2), such as the ones illustrated in Figure S3a-e, exhibit an apex ROC saturation around 1.9 μm after being annealed for 1 h at 1200 °C, as mentioned in the main text. Increasing the AR to 3–6 slows the growth of the apex ROC during prolonged annealing at 1200 °C, reaching approximately 1 μm after 1 h. However, the typical blunting of the apex caused by surface diffusion of Si adatoms is accompanied by the formation of a droplet-like profile characterized by a pinch-off region, as surface diffusion minimizes surface energy in accordance with the Plateau–Rayleigh instability[55]. This phenomenon is illustrated in Figure S3f-j, where a Si probe fabricated by focused ion beam (FIB) milling with an AR of ~4 develops a distinct pinch-off point after 65 min of annealing at 1200 °C. For probes annealed in the inverted (apex pointing down) orientation, the pinch-off occurs directly beneath the apex (Figure S3j), in contrast to upright Si pillars where it emerges near the base of the structure[55] . This pinch-off behaviour imposes a maximum viable annealing duration: beyond this limit, continued surface diffusion can detach the apex from the probe, leaving behind debris or irregular features on the newly formed surface. As a result, the new apex is neither atomically clean nor flat, and additional annealing would be required to restore its morphology—at the risk of triggering another detachment event. Further increasing the AR accelerates this pinch-off process, as evidenced by a probe with an AR of ~6 that developed a pinch-off after only 30 min of annealing at 1200 °C. In the extreme case of an AR of ~ 10 (Figure S3k-o), the probe undergoes accelerated shrinkage even at 1000 °C (see Figure S3m), attributed to frequent pinch-off events arising from the narrowing of the probe diameter, consistent with the Plateau-Rayleigh instability. At 1100 °C and 1200 °C, these high-AR probes, initially ~5 µm in length, collapse into the substrate within 3 min and 1 min, respectively. Based on these findings, we fabricate Si probes with an AR of approximately 1.5 to ensure reproducible formation of atomically clean and flat apexes, while pursuing the integration of low-temperature surface-cleaning methods into our probe preparation process to expand control over the build-site dimensions and broaden the applicability of mechanosynthesis.

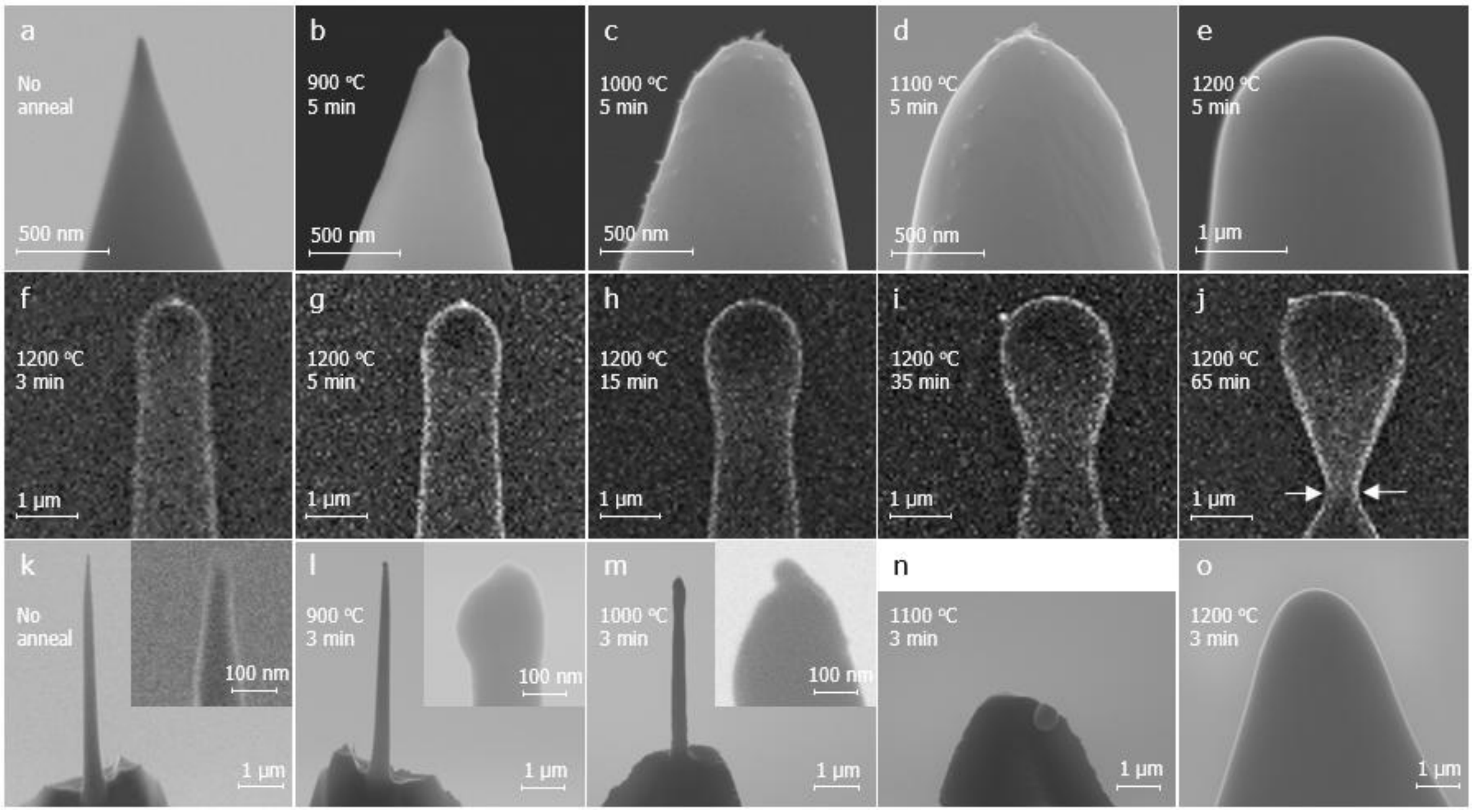

*Figure S3:* ***Annealing Mechanics of Si Probes with varying Aspect Ratio (AR)****: a-e) SEM images of five distinct Si probes with low AR (roughly <2) after an anneal treatment, starting with no treatment applied in (a), followed by 5 min anneals from 900 °C to 1200 °C in increments of 100 °C in (b-e). A smooth and debris-free apex is observed only in the case of (e). f-j) SEM images of a single Si probe with high AR (~3 to 6) after increasingly longer anneals at 1200 °C, starting at 3 min and ending at 65 min. Over the course of the five anneals, the apex develops a teardrop shape, resulting in a narrow pinch-off region below the apex (white arrows in j) which culminates in the physical detachment of the apex from the probe upon further annealing. k-o) SEM images of five distinct Si probes with extreme AR (>9) after an anneal treatment, starting with no treatment applied in (k), followed by 3 min anneals from 900 °C to 1200 °C in increments of 100 °C in (l-o). High magnification SEM images of the apex are shown in the insets of (k-m). The extreme AR probes are fabricated using similar probes as in (a) and Focused-Ion Beam milling. These extreme AR probes shrink at an accelerated rate and therefore cannot withstand high temperature annealing (>1100 °C), required for a clean Si surface, as seen in (o).*

# 5. Role of Relative Probe Tilt in Inverted-Mode STM

In Inverted Mode STM, surface-bound molecules on a flat Si sample act as individual imaging tips that generate reflected probe images (RPIs) of the Si probe apex microfabricated on a Si chip, referred to as SPC. The appearance of these RPIs varies strongly with the height, imaging orbital, and orientation of the imaging molecule. The height and imaging orbital of the molecule are intrinsic to its chemical structure, although the latter can be modified in situ, as described in *Mechanically Controlled Chemical Reaction* in the main text. In contrast, the orientation of the molecule relative to the topmost Si(100) terrace of the probe apex is dictated by the molecule's binding configuration – governed by surface chemistry – and by the relative tilt angle between the Si(100) surfaces of the SPC and the Si sample. The latter is required to be small to produce an RPI containing a fully visible topmost terrace.

In practice, small but non-zero tilt angles between the SPC and the Si sample – along X and/or Y directions – routinely occur after mounting these components into the STM system and typically range within ±1°. The resulting RPIs can be categorized by the visibility of the topmost terrace. For instance, in Figure S4b-d, EAO-GeC2I molecules (heights of 300-350pm) is utilized to obtain three experimental RPIs in which the topmost terrace appears fully visible at tilt angles ($\theta_X$, $\theta_Y$) of (0.23°, 0.02°), partially visible at (0.7°, 0.05°), and inaccessible at (0.9°, 0.57°), respectively. Although the precise tilt angle threshold between these regimes depend on the molecule height, the overall trend remains consistent: as the tilt angle increases, the visibility of the topmost terrace decreases and progressively lower terraces emerge within the RPI.

A corresponding series of simulated RPIs can be generated by applying geometric tilt transformations to a conventional STM image of a Si probe apex (Figure S4a). The simulation involves three steps: (1) linearly adjusting the Z values of the STM data along the X and Y axes according to the specified tilt angles $\theta_X$ and $\theta_Y$; (2) normalizing the highest Z value to zero, and (3) truncating the data such that the magnitude of the lowest Z value does not exceed the height of the imaging molecule. The simulated RPIs (Figure S4e-g) reproduce the experimental images in Figure S4b-d, confirming that the tilt angle predominantly determines the topmost terrace visibility.

The constant and intrinsic tilt between the sample and the probe results from the manufacturing tolerance in the STM head. One can adjust the angle of the sample or probe with calibrated spacers, but the process is slow. Motorized goniometers or similar alignment modules can be integrated in the STM head for in-situ control of the tilt.

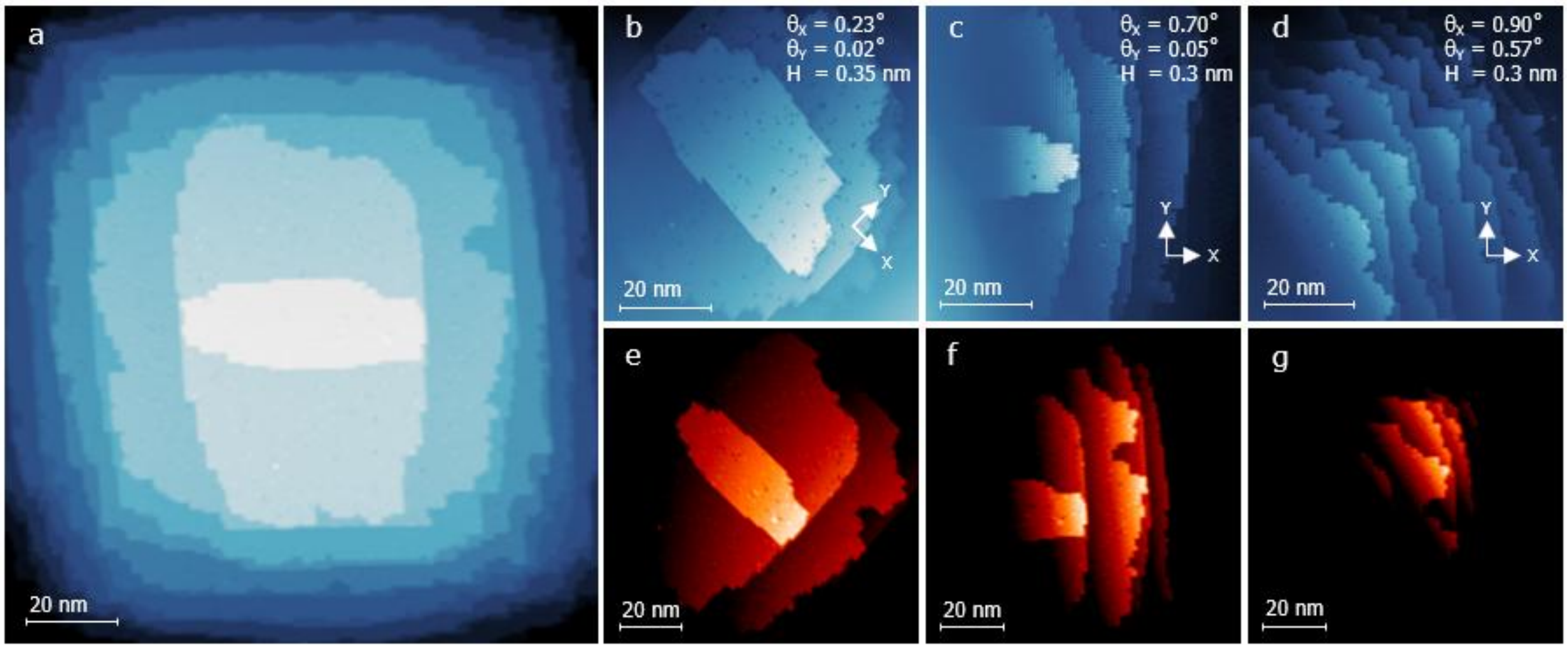

*Figure S4:* ***Impact of Relative Probe Tilt on RPI*** *: a) Conventional mode STM image of an annealed Si probe apex obtained using a sharp metal tip. b-d) Inverted mode STM images showing RPIs of three distinct Si probe apices using EAO-GeC2I as the imaging molecular tool. The relative tilt between the Si probe apex and the EAO-GeC2I molecule, denoted in each image by ($\theta_X$, $\theta_Y$), increases gradually from low (b) to moderate (c) and finally to high (d). e-f) Numerically modified versions of (a) meant to resemble the experimental RPIs in (b-d), respectively. The tilt angles ($\theta_X$, $\theta_Y$) and the EAO-GeC2I molecular height (H) shown in (b-d) were input parameters to simulate RPIs in (e-g). The simulated RPI in (e) was also rotated to resemble more closely the experimental data in (b).*

# 6. Explorations of Using Inverted-Mode STM for In-Situ Probe Preparation Methods

In addition to the annealable SPCs developed and demonstrated in the main text, several alternative in-situ methods of producing and restoring planar, crystalline probe apexes were also explored, with inverted-mode STM used to assess their success. These alternatives can be performed without significant hardware modification, albeit with less reproducibility of the final probe structure. *In situ* crystalline probe preparation is also another demonstration of the utility of molecular tools as a readily available reference point to assess probe shape and cleanliness, whether for inverted mode SPM or probe conditioning in general. Some highlighted results of the in-situ conditioning of tungsten (W) and silicon (Si) tips are described below, though we anticipate analogous outcomes for other probe materials as well.

## *In-Situ* Field Evaporation and Extreme Conditioning of Tungsten Probes

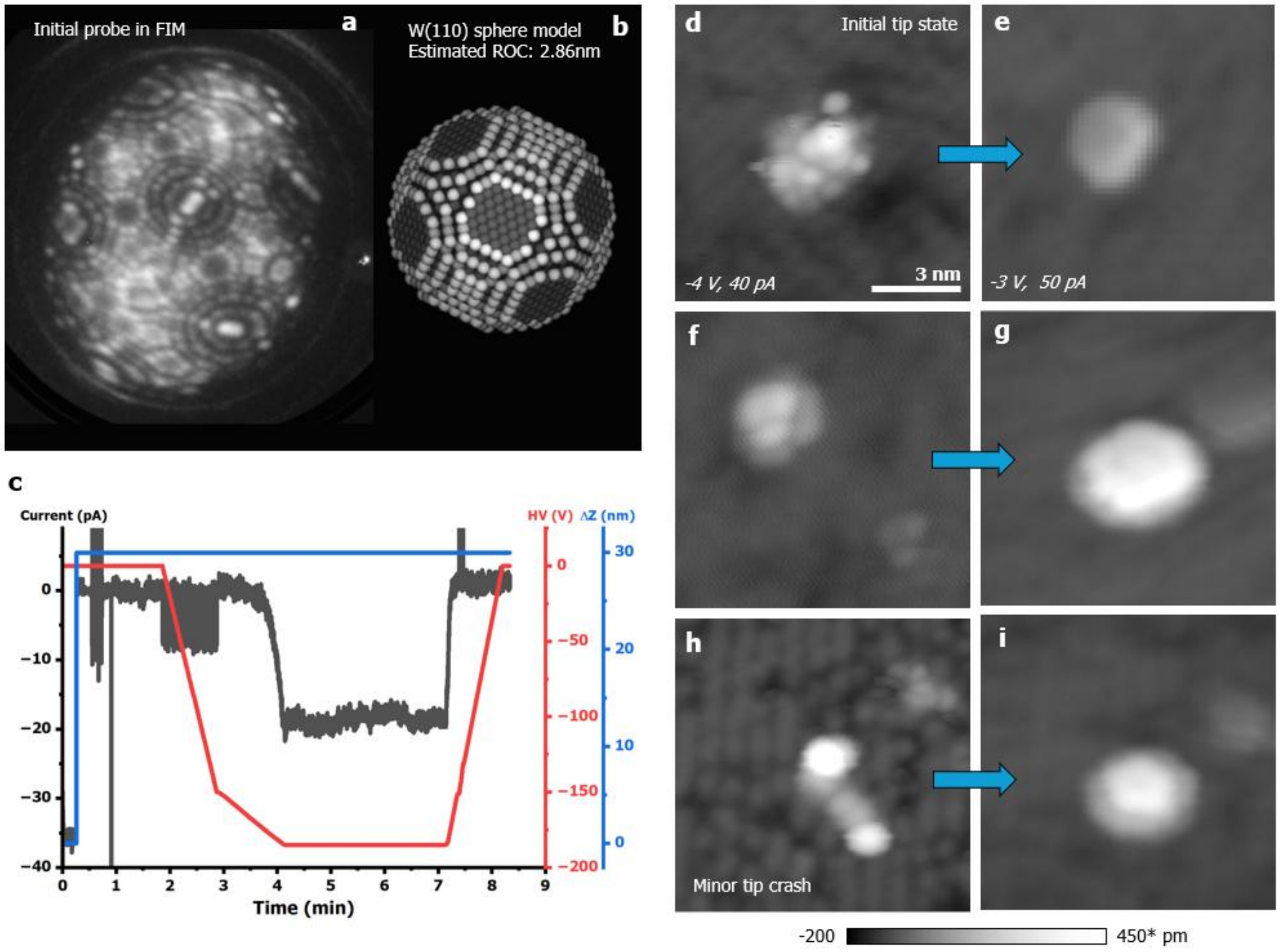

*Figure S5: Demonstration of inverted mode STM imaging used for W(110) probes. (a/b) Preliminary FIM-based preparation of an as-etched W probe, including an estimate of effective radius of curvature based on a W(110) spherical model. (c) A typical high-voltage field-evaporation procedure used to clean probes* in situ. *(d-i) The typical life-cycle of a W(110) probe with inverted-mode STM, images generally taken at -3 V, 50 pA except where noted. (d) Initial W probe state as imaged by a molecule after transfer from FIM to UHV via air. (e) The result of multiple iterations of the procedure noted in (c), yielding an atomically-clean W(110) facet. (f/g) Later probe states achieved with a W(110) trimer apex and another, atomically-flat*

*probe apex. (h/i) In situ recovery of the same probe from a minor probe crash via the procedure in (c). *The relative Z height scale maximum for sub-panel (i) is 525 pm.*

An example of the utility of inverted-mode imaging of the STM probe was demonstrated using more traditionally-prepared W tips. Following the procedure outlined in *Sample Preparation and STM Measurements*, atomically-clean W(110) probes were finalized and confirmed via FIM before rapid transfer through air (Figure S5a/b).

*In-situ* preparation methods were explored for tungsten probes, targeting atomically clean terraces, with some success. For initial *in situ* probe preparation and reconditioning, a high-voltage field evaporation process was employed with a 56 kΩ protection resistor between the tunneling current output and the STM pre-amplifier to protect it from inadvertent shorting at >10 V biases (Figure S5c). The probe was retracted by 10-50 nm from the tunneling setpoint, and the bias was ramped upward, typically above 100 V, until a small, consistent field-emission current (<50 pA) was measured. This current was maintained for several minutes before returning to tunneling conditions and inspecting the probe state. Discontinuous steps in current could be monitored in some cases, highlighting local changes in apex geometry during evaporation. Figure S5d-i demonstrates various W probe apexes as imaged by surface-bound molecules, including an initially disordered state (presumably with an adlayer of contaminants) which could be cleaned to an atomically-clean W(110) facet (Figure S5d/e), and recovery from a minor tip crash into the Si(100) surface (Figure S5h/i). Such atomically well-defined probe apexes may be particularly useful for certain types of spectroscopic studies, where the composition of the probe apex is especially impactful.

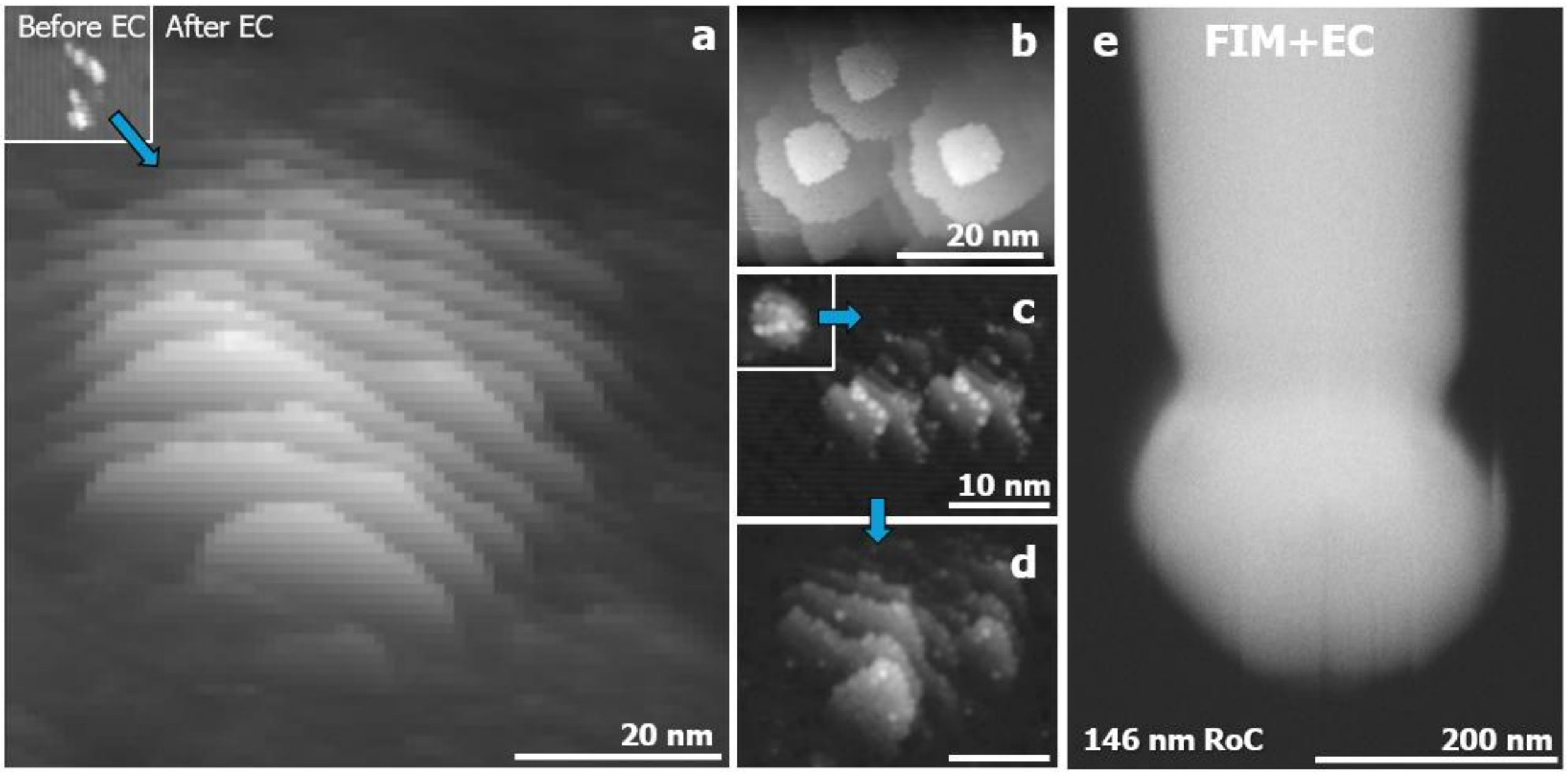

*Figure S6: Results of "extreme conditioning" (EC) of W(110) probes as evidenced by molecular tool-based reflected probe images (RPIs). (a) A demonstration of a sharp, jagged RPI (inset, same scale) revealing multiple W(110) terraces accessible to the molecular tools (main panel) after EC. (b) An RPI containing the apex terrace of the W(110) sphere after EC. (c/d) A W(110) probe with sequential extreme conditioning events starting from an initial, compact RPI (inset). (e) Typical SEM image of a W probe after EC. Accelerating voltage = 5 kV, secondary electron detection, vacuum < 5E-6 mB.*

In addition to field evaporation, another procedure, which we call "extreme conditioning" (EC) was found to produce atomically clean W(110) apexes with larger radiuses of curvature. This procedure is similar to the one outlined in Figure S5, but with higher bias and/or less Z separation between the tip and sample. In these cases, discontinuous drops in current were observed along with mesoscopic changes to apex. In one step, RPIs could be converted from disordered, sharp apexes to crystalline, terraced W(110) planes (Figure S6a and inset). Such EC-treated probes are much flatter and broader, leading to changes in the tip-induced band bending[68,69], but present a unique tip-sample configuration for certain types of spectroscopic studies, especially of other, similarly sharp molecules. In those cases, imaging of the substrate may be of less utility than a well-ordered and visible metallic tip state from inverted-mode imaging. For example, being able to access the top terrace of the W(110) sphere after EC treatment gives a well-defined tip geometry, even if individual W atoms are not readily resolvable with the given molecular tools[70–72].

EC treatments could also be applied multiple times to reclean contaminated apexes, though the radius of curvature tends grow with each treatment until the electric field is insufficient to provoke further reforming of the W tip (Figure S6c/d). Subsequent SEM analysis of these EC-treated probes revealed apparent melting at the apex with a globular or spherical termination (Figure S6e), suggesting a less-controlled variety of current-driven apex annealing than either SPC annealing or more typical field evaporation-based probe reshaping. Recrystallization is thought to be influenced both by the temperature of the system, as well as the influence of the local electric field, which may preferentially bias W atom migration as the probe stabilizes[70–72], though further studies are warranted. These tips, once blunted, cannot be sharpened with FIM-like tools, and must be re-etched to present a fresh, sharp apex with the appropriate field-enhancement to be used in this way. In the most extreme cases, the mesoscopic shape of the W apex could be changed as well, leading to curling of the W probe apex away from the tunnel junction, but with a clean, spherical apex structure (so, not the result of mechanical contact). Curled probes of this type would need to be more thoroughly etched or replaced altogether, similar to blunted ones.

## In-situ Preparation of Crystalline Si(100) Probes

Similarly to tungsten, an *in-situ* probe conditioning protocol was developed to create conductive, crystalline, flat silicon probe apexes, suitable for scanning and localized chemical interactions (i.e. mechanosynthesis) at 4K. Some literature precedent exists for *in-situ* growth of Si nanocolumn structures on an STM tip[56,58]. It has also been shown that large bias pulses can sometimes result in a tip that is coated in sample material that exhibits crystallinity, hypothesized to be a result of a local cleavage of the surface with the aid of the STM tip[57]. These early demonstrations primarily employed metal tips coated in silicon via pulsing, on Si(111) samples, and focused on growing nanoneedle structures on the tip apex with little focus on apex control of the nanoneedle structure. The *in-situ* probe conditioning described herein was an attempt to expand on these early works.

The tips were assembled by silver epoxying a commercially available highly n-type doped silicon AFM cantilever (Nanosensors PPP-NCHR) onto a 150 um diameter tungsten wire. The tip assembly was then moved into the STM head, where we performed several rounds of field emission to remove the native oxide layer, until stable tunneling of the silicon probe over the sample was achieved (the sample used was a Si(100)-2x1 surface, deposited with EAOGe-C2I molecules).

Many approaches to modifying the probe apex were explored, mostly involving dynamically adjusting the bias and probe height as a function of time. All such approaches to *in-situ* probe conditioning had significant randomness in the outcomes. Changes induced to the probe included not only its shape and atomic configuration, but also its conductivity. Nonetheless, with many trials we found that we could repeatedly prepare conductive crystalline silicon probes. Two processes, which we call “contact bridging” and “positive bias plunges”, were found to be useful, and they will be described below. The data shown in this section provides an example of a case in which the combination of these two processes led to a clean apex, but this represents a best-case scenario, and most trials did not lead to a clean, well-ordered apex.

Contact bridging, an example of which is shown in Figure S7a-d, is the first step and it is primarily intended to flatten the probe and improve its conductivity. This involves first pushing the probe into the sample by 1-10 nm at 0V to make a contact, starting from the height defined by the tunneling setpoint. While still in contact with the surface, we ramp the bias from 0V up to a maximum value on the order of ±20-40 V. As the bias is ramped up, the current channel is monitored – if an ohmic contact between probe and sample has been made, the current will linearly increase as the bias is ramped; we typically observe current values on the order of hundreds of uA passing through the probe-sample contact region at maximum voltage, well above tunneling conditions. A 56 kΩ protection resistor was inserted between the tunneling current output and the STM pre-amplifier to protect it from inadvertent shorting. While maintaining the high bias, the probe is then slowly retracted, at a rate on the order of 1 nm/s for at least 20 nm. We observe the high non-zero current is maintained for some length of the extrusion process and hypothesize that this non-zero current is indicative of a nanoscale structure (which we refer to as a “contact bridge”) having formed between the probe and sample that is slowly elongated while the non-zero current is measured. In the example of Figure S7a, the high current signal was measured for the first ~15 nm of retraction, at which point it quickly dropped to zero. This is consistent with previous studies which demonstrated growth of nanowires on a probe apex, using a similar methodology[73] and more recently with metal tips and surfaces[74]. After this extrusion step, the bias is reduced to the tunneling bias, and the probe is quickly withdrawn from the surface. We then collect an RPI of the resulting probe apex to determine if the process has produced the desired flattened, crystalline probe structure. Typically, the flattening process involves a series of contact bridging attempts, usually of increasing bias and poke depth, until a crystalline apex structure is observed.

Once the probe apex shows signs of having a crystalline terrace, positive bias plunges, an example of which is shown in Figure S7e-h, are employed to remove Si adatoms from the apex, to reveal the larger terrace underneath. A positive bias plunge starts by ramping up the bias to 7-9 V at about 1 V/s, with feedback off, then moving the probe toward the surface by about 600-800 pm at a rate of 200 pm/s. The probe is held in this position for 2-3 seconds, then withdrawn about a nanometer above the initial starting position over 2-3 seconds. The applied bias is ramped back to the initial condition, and an RPI of the resulting probe apex is captured to assess the success of removal of surface adatoms.

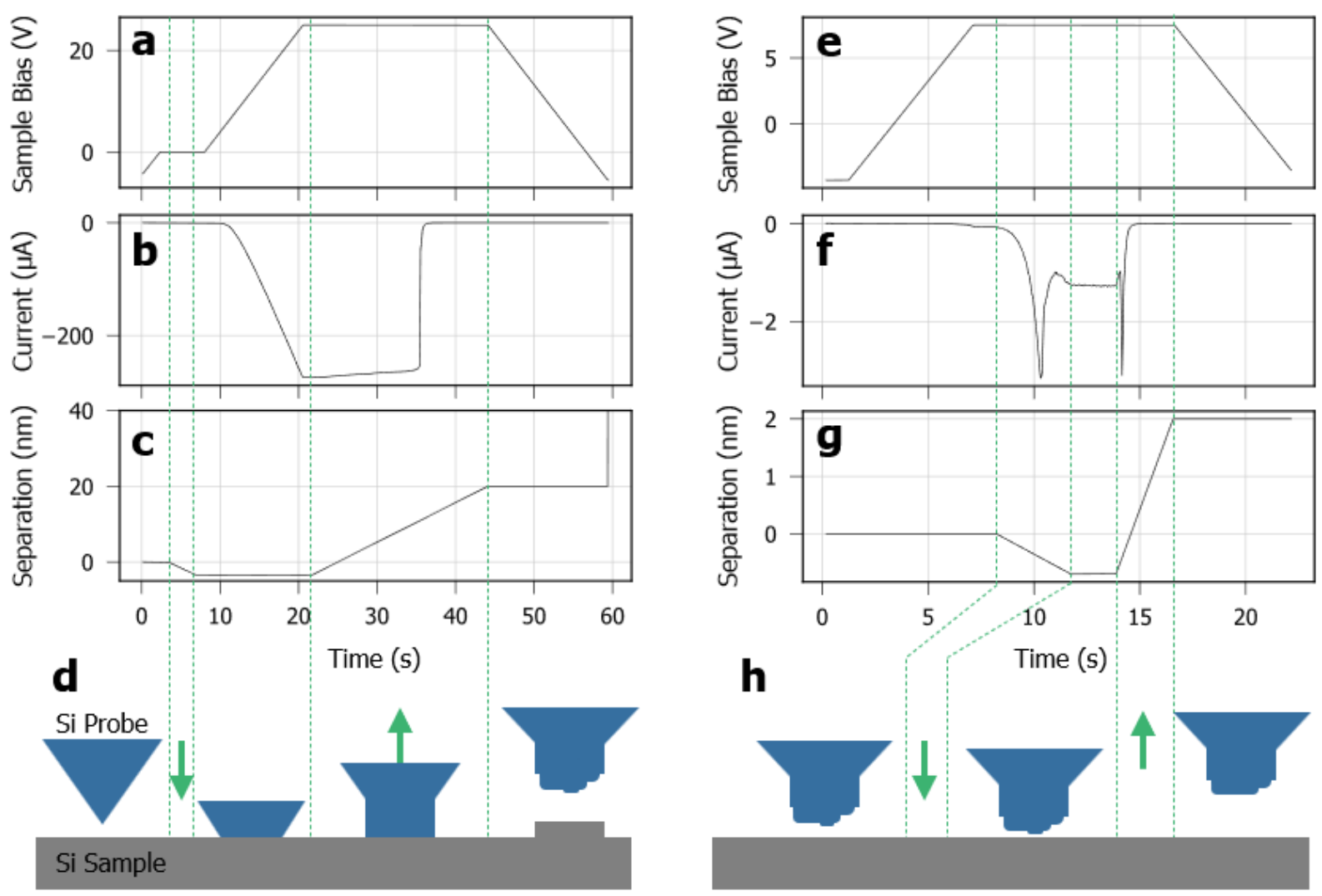

*Figure S7:* ***Contact bridging and positive bias plunges.*** *a) Sample bias, b) tunneling current, and c) probe-sample separation relative to the starting separation, as a function of time during a contact-bridging procedure. d) Schematic illustrating the intended effect on the probe condition during contact bridging: from an initial height, the probe is pushed into the sample and bias is ramped up; with bias maintained, the probe retracts with the intent to form a crystalline bridge connecting probe and sample; after retraction, the cleaved probe apex ideally shows indications of crystallinity and atomic terraces. e-g) The same three quantities as a function of time, but for a positive bias plunge. g) Schematic illustrating the intended effect on the probe condition during positive bias plunges: the probe moves toward that sample and bias is ramped up; then the probe retracts under high bias. This process is meant to remove adatoms and reveal a clean topmost atomic terrace. Note that the data shown corresponds to the conditioning steps utilized to create the crystalline probe apex shown in Figure S8.*

The probe apex that was made using the contact bridging procedure of Figure S7a-c is shown in Figure S8a. Positive bias plunges were repeated several times to remove the surface atoms, revealing the full second terrace below (Figure S8b-i). The final positive bias plunge (performed between panels h and i) is the one shown in Figure S7e-g. Inverted-mode imaging is used to characterize the probe apex after each conditioning step. Panel j shows a larger area, where several RPIs are visible.

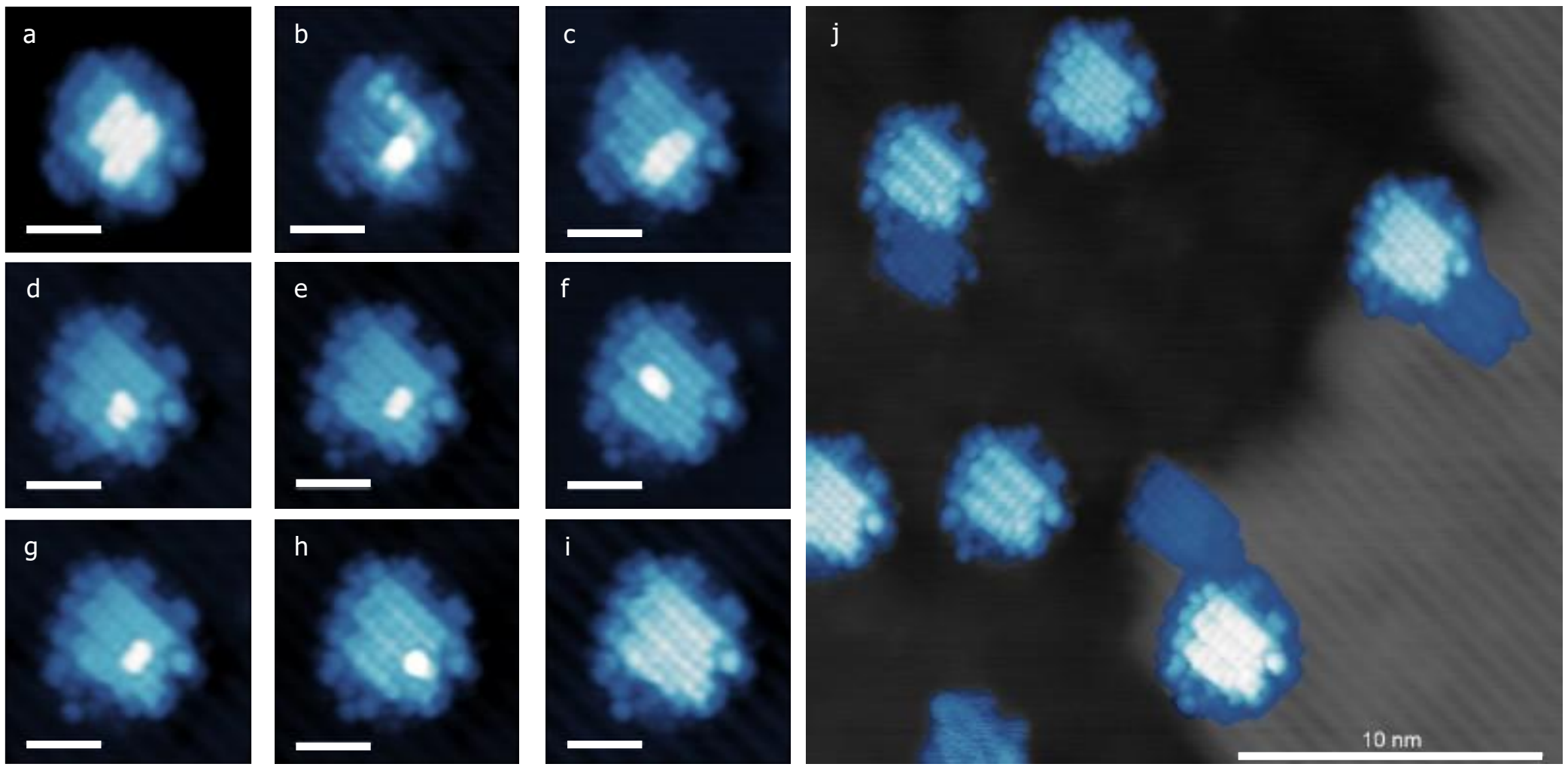

*Figure S8: Set of inverted mode STM scans of the silicon probe apex using EAOGe-C2I molecular tool. a) Silicon probe apex after a contact bridging attempt that reveals a crystalline 100 surface reconfiguration. b-i) A sequential series of scans were taken after each positive bias plunge was performed to remove silicon atoms on the apex until the terrace below was revealed. The scale bars represent 2nm. Image j) shows a 25nm X 25nm region where several molecular tools are imaging the same final probe apex, showing several different imaging qualities. All images captured at $V_S$ = -4.25V, $I_T$ = 50 pA.*

When the processes described so far failed to produce the desired outcome and/or when probe conductivity was lost, we employed more aggressive forms of tip conditioning, like pushing the probe into the sample with or without bias or applying bias pulses. The goal was to obtain a probe that was tunneling first, then use contact bridging or positive bias plunges to improve conductivity, flatten the apex, and achieve crystallinity.

It is noteworthy that the orientation of the Si(100) domain on the probe apex in Figure S8 matches the orientation of the Si(100) sample surface. This was observed in nearly all cases of crystalline probe apexes produced *in situ*. This leads us to believe that the bridging process grows epitaxially from the surface, and when the bridge is presumably cleaved, it exposes a (100) face that matches the surface in tilt and rotation. This phenomenon was only observed on Si(100) and attempts to reproduce it on Si(111) were inconclusive.

Finally, it should be noted that the *in-situ* Si probe preparation process that has been developed here has produced several probes with crystalline apexes, but the contact bridging process is highly stochastic. To obtain a crystalline Si probe apex can take anywhere from a few hours to a month to complete, and there is little control of the final probe apex structure. The stochastic process of contact bridging, lack of control in final probe structure, and time requirements further motivated the development of the SPC probes to ensure a relatively quick and reproducible method for generating crystalline probe apexes for inverted mode STM.

# 7. Further SPM Characterization of EAOGe-C2I

Figure S9a shows an STM image of a Si(100) sample deposited with EAOGe-C2I at a typical coverage for inverted-mode STM, but in this case characterized by a sharp tungsten probe. In the image, there are two tall round protrusions with heights around 550 pm, indicated by white arrows. Additionally, there are several shorter features with heights in the range from 200 to 300 pm. Sampling the molecular heights of 48 features across several images taken with a sharp probe yields an average of 252±153 pm, with a range up to 600 pm but a large number of short features (~100-150 pm in height) that could be other products (e.g. molecular tools in different orientations, that decomposed during adsorption, or other contaminants from the deposition). Figure S9b shows a similarly prepared sample, but scanned by an annealed SPC. The density of RPIs is similar to the density of the tallest protrusions in the sharp-probe images. Sampling 48 RPIs, we calculate an average height of 288±72 pm, but with a lower maximum than sharp-probe imaging of ~400pm. Scanning with an SPC acts as a filter for EAOGe-C2I tools with the iodoalkyne oriented upright, in that the large majority of RPIs seen in large-area images can be deiodinated and used for mechanosynthesis, as discussed later in this section. In contrast, in sharp-probe imaging the majority of features are short and probably not upright-C2I tools.

Carbon-iodine bonds are known to undergo homolytic scission under UV-illumination. Figure S9c shows an STM image of a sample after exposure to UV light, taken with an annealed SPC. After UV illumination, there are two distinct populations of RPIs. There are RPIs of moderate heights, indicated by black arrows, which match the RPIs seen in non-illuminated samples like the one shown in Figure S9b. Additionally, there is a second set of RPIs that are noticeably taller and that provide qualitatively different images of the probe apex (with lower resolution, like in Figure 4), indicated by white arrows. For much greater UV-exposure, nearly all molecules are found to be in the taller state. This observation strongly suggests that the bright RPIs are produced by de-iodinated molecules. The UV exposure used for the sample in Figure S9c was chosen to achieve a similar number of molecules in each state, and the proportion of bright RPIs on this sample was about 43%. Notably, some scanning conditions can induce changes to the shorter RPIs, making them taller and indistinguishable from the set of RPIs created by UV-activated molecules, indicating that molecules can also be deiodinated by the probe, *in situ*.

Figure S9d and e show a region of an SPC apex imaged in inverted mode before and after tip-induced deiodination. The initial image (Figure S9d) shows a clean Si(100)-2x1 surface. The imager resolves dimers, albeit with an asymmetric character that is frequently seen with EAOGe-C2I. We have found that positive sample biases (around 4 V) have a high likelihood of inducing deiodination of the molecule. Figure S9e shows the same region of the SPC imaged after such a procedure, performed with the probe located at the center of the image. The entire image is shifted in height by about 330 pm and the imaging molecule no longer resolves the individual dimers. Additionally, a new protrusion that switches back and forth on a single dimer appears on the SPC apex at the location where the deiodination was performed. We interpret this new feature as a single iodine atom, the one that was removed from the imaging molecule.

The observations made with UV-illuminated samples, and their correspondence with probe-induced changes to the imaging molecules, provide a strong base of evidence for our interpretation of the molecule's orientation and composition. In particular, they suggest that many molecules adsorb with their iodoalkynes oriented upward (away from the surface to which the molecule is bound), and that the capping iodine atom can be removed either by UV light, or by probe-induced processes.

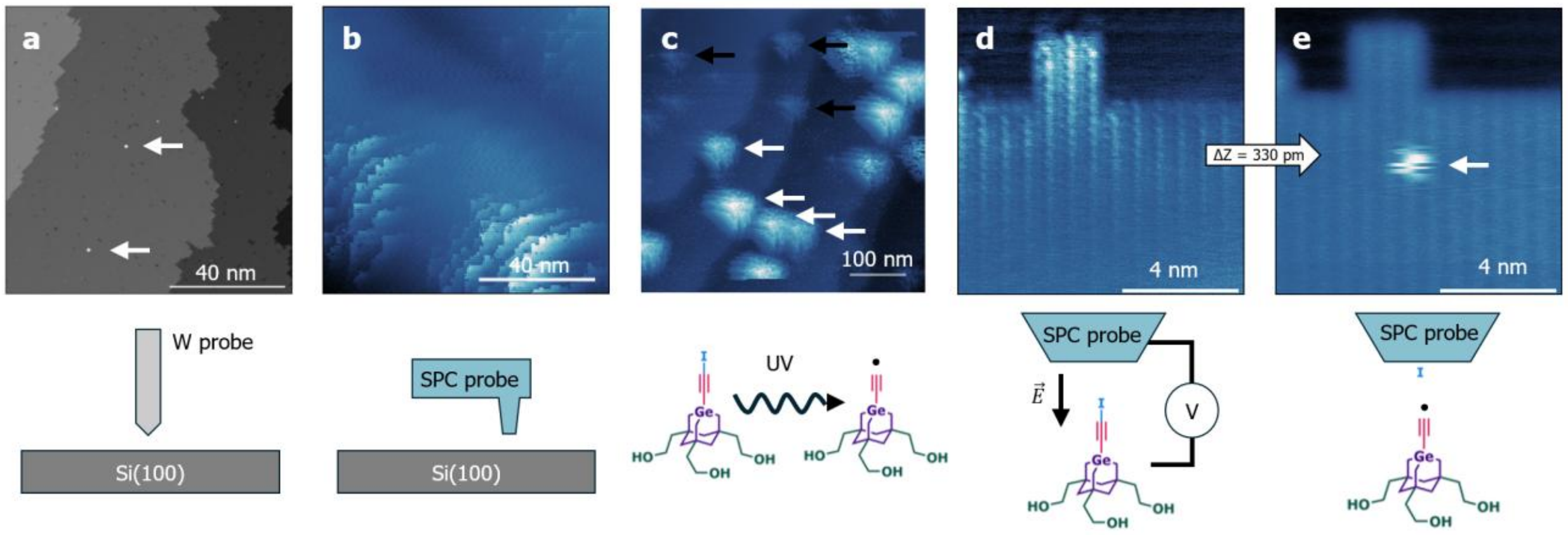

*Figure S9:* ***Characterization of EAOGe-C2I by STM.*** *a) STM image of a Si(100)-2x1 sample with a sparse coverage of EAOGe-C2I, taken with a sharp tungsten probe in "conventional mode" [$V_S$ = -2 V, $I_T$ = 30 pA]. White arrows point to tall protrusions with heights around 550 pm. Shorter protrusions, with heights from 200 to 300 pm, are also visible. b) Inverted-mode STM image, taken with an annealed SPC, of a sample prepared by the same method as for (a) and with a similar coverage [$V_S$ = -3.5 V, $I_T$ = 30 pA]. Multiple RPIs can be seen, each of which is produced by a single protrusion on the sample. c) Overview STM image of a sample prepared by the same method [$V_S$ = -3.5 V, $I_T$ = 4 pA], but exposed UV light from an LED source centered at 265 nm with an intensity on the order of 10 W/cm2, for a total exposure time of 15s. Whereas some RPIs, indicated with black arrows, are of a similar height to the ones shown in (b), others (indicated by white arrows) are significantly taller. d) Small region of an SPC apex imaged in inverted-mode STM using an iodinated molecule as the imager [-3.8 V, 10 pA]. e) Same region of the apex as (d), but after bias-induced deiodination. The average image height increased by 330 pm, and a new protrusion, attributed to an iodine atom and indicated by the white arrow, can be seen. [-4 V, 5 pA].*

# 8. More information on H-passivation of SPCs

This section summarizes the procedure for hydrogen passivation of an SPC inside the UHV probe preparation chamber. Hydrogen passivation of silicon is routinely done at various labs and our procedure generally follows accepted methodology[75,76].

In order to passivate the probe, two conditions must exist:

1. The probe preparation chamber must be plumbed for leaking in high purity hydrogen gas, and fitted with a thermal gas cracker to crack the hydrogen gas into atomic hydrogen;
2. The annealing recipe for passivating the specific SPC probe must be predetermined via a probe annealing calibration step.

The plumbing setup is shown in Figure S10, where “UHV” represents the probe preparation chamber, whose pressure is maintained on the order of 1e-10 mbar, GH2 represents a tank of ultra-high purity H2 gas (99.999%) which is leaked into the chamber via the leak valve (LV).

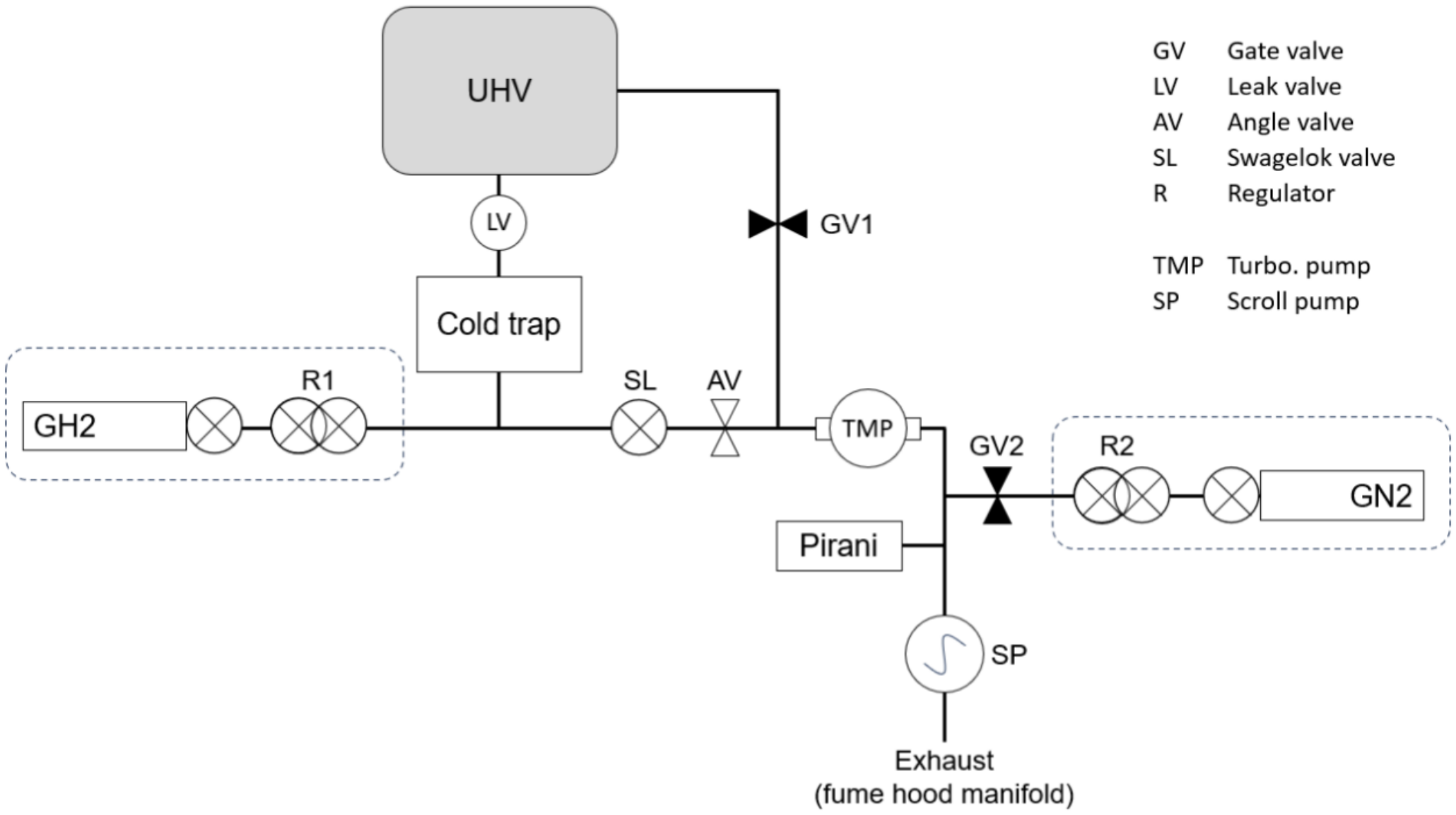

*Figure S10: Hydrogen gas plumbing configuration for passivating an SPC. The SPC is annealed in the “UHV” chamber. Under typical operation, the H2 lines are kept under vacuum by a turbo pump, TMP. During hydrogen passivation, the TMP is shut off and closed to the H2 lines via the Swagelok valve, SL, and high purity H2 gas (99.999%) is introduced into the chamber via the leak valve, LV.*

After installation and initial bake-out, ultra-high purity H2 gas (99.999%) is flushed through the plumbing several times to remove any undesired species in the plumbing. Subsequently, the H2 lines are kept pumped on with a molecular turbo pump until immediately before a passivation run, to maintain cleanliness in the lines. A liquid nitrogen bath (i.e. a cold trap) can be used to further reduce the partial pressure of contaminants within the plumbing by freezing out any molecules other than H2 during a passivation run.

During passivation, the SPC is placed in the probe preparation chamber and H2 gas is leaked into the chamber, to a pressure of 1e-6 mbar. DC current is passed through a homemade tungsten (W) filament thermal gas cracker located inside the chamber, heating it to a temperature of 1800°C and converting H2 gas into atomic hydrogen. The SPC is then annealed in the presence of the gas, allowing the probe chip to become passivated.

In general, the annealing recipe for passivation consists of a quick “flash” step, followed by a long “hold” step. The SPC chip is heated to 1200°C for 3 seconds to remove any contaminants on the surface, followed by a 330°C hold for 5 minutes, all in the presence of atomic hydrogen. The long hold promotes the formation of the desired H:Si(100)-2x1 surface reconstruction. As mentioned previously, annealing of the SPC is done via direct current annealing. Prior to any passivation run, the annealing recipe must be predetermined using a calibration run to ensure the probe apex reaches the desired temperatures. Once passivated, the probe is quickly transferred into the STM head.

The figure below shows images of the same SPC taken directly before and immediately after passivation. A large (400 nm x 400 nm) overview image is always taken (S11 a and b), to find suitable molecules for use in apex characterization and mechanosynthesis. The insets show zoomed in images of the probe apex (60 nm x 60 nm), captured with the molecular tools shown outlined in white. After the quality of the bare SPC apex is confirmed (panel c), the probe is withdrawn from the head, passivated using the procedure described above, and then reinserted into the head with the same sample.

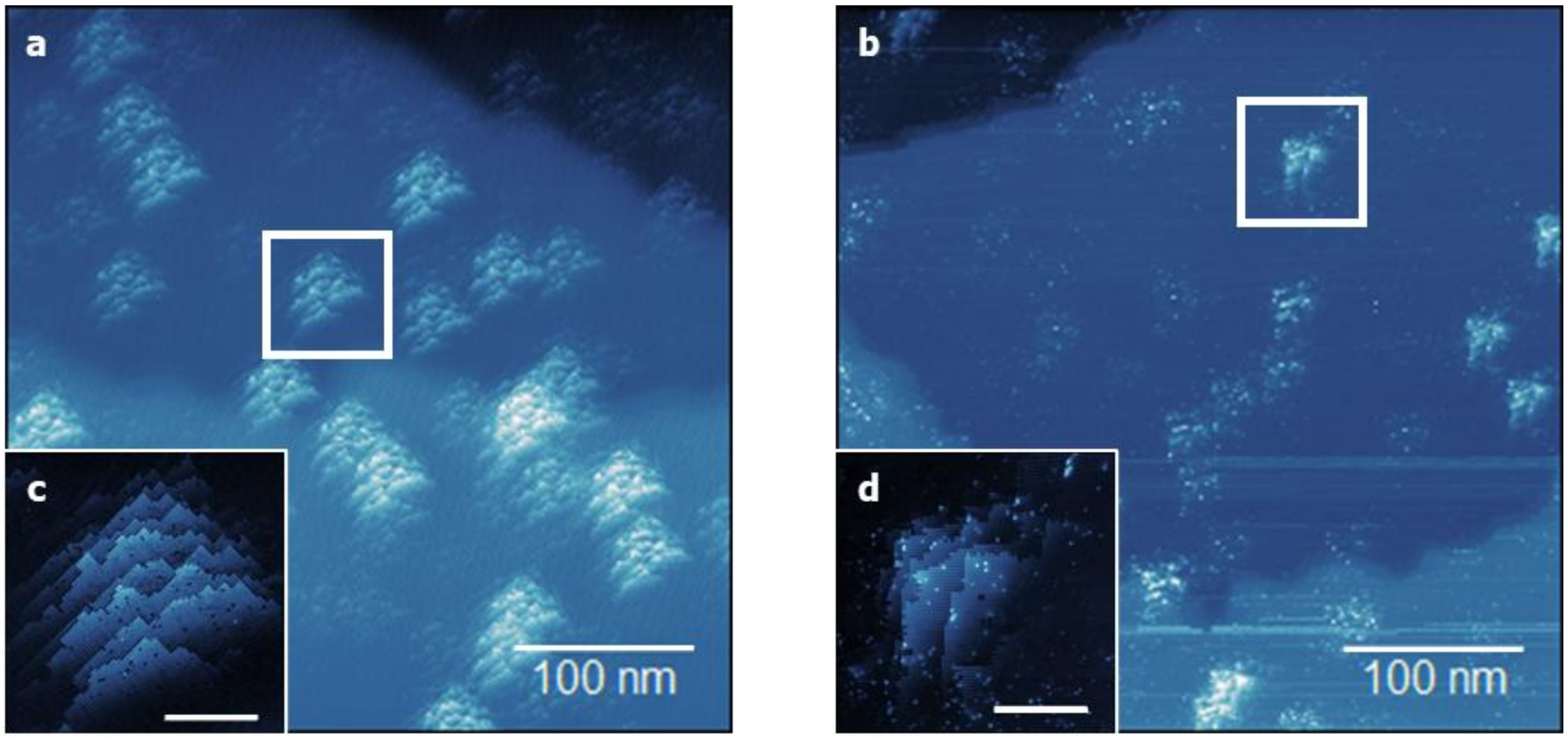

*Figure S11: Figure showing the same SPC before and after H-passivation using the procedure described in this section. Overview scans (a and b) each show many RPIs of the SPC. The insets (c and d) show a full RPI image of the specific RPI outlined in the white box. The inset scale bars are 20nm. Images captured at $V_S$ = 3.2V, $I_T$ = 10 pA.*

Note that because the same sample is used in both cases, one immediate observation that can be made from this data is that only a subset of the molecules on the surface image the H-passivated SPC well. It is yet unclear why this is the case. Note that there is a 45° rotation of the scan frame applied to the images taken after probe passivation, to orient the probe dimers to the scan frame.

# 9. Mechanical Constraints on Reagent Configurations

The chemical reaction discussed in the main text can be viewed through the lens of a Potential Energy Surface (PES), as shown in Figure S12. The potential energy of the system is shown as a function of the silicon-to-hydrogen distance and the carbon-to-hydrogen distance, schematically illustrated in the callout to the left. The PES has a valley on the left side, which represents the case where the hydrogen atom is bonded to a surface silicon atom of the H:Si(100) probe. There is also a deeper, energetically favored valley along the bottom of the PES, which corresponds to the case where the hydrogen atom is bonded to the carbon atom of the molecule. The separation between the silicon atom on the probe side and the uppermost carbon atom on the sample side can be adjusted, to a good approximation, by the STM's scanner along its *z*-axis. This means that at constant *z*, the system is constrained to diagonal lines of constant separation, and changes along those lines correspond to the movement of the hydrogen atom away from one side and toward the other.

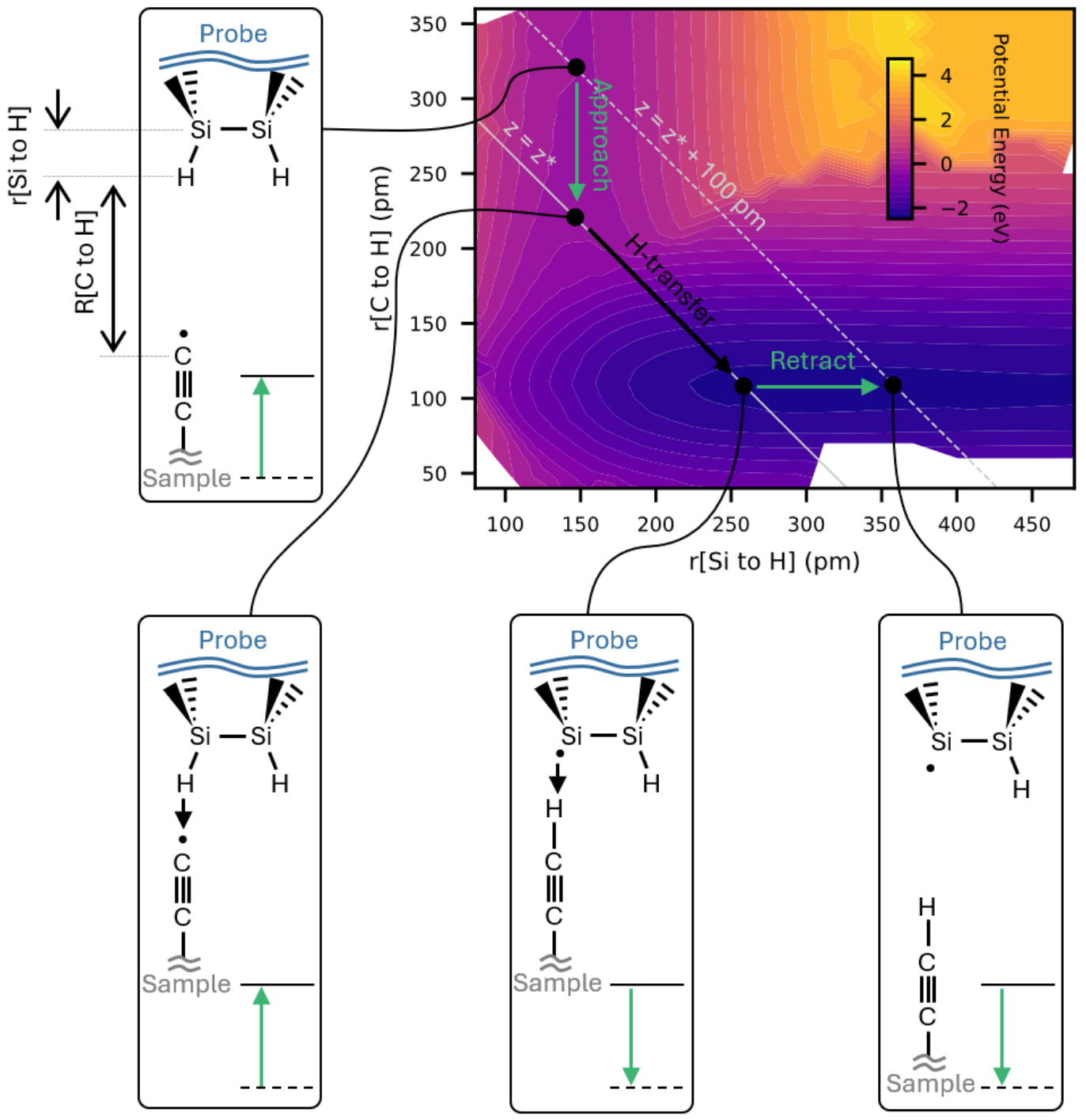

*Figure S12:* ***Potential Energy Surface (PES) for H-transfer:*** *Potential energy of the chemical system as a function of two reaction coordinates shown in the callout on the left: the Si-to-H distance on the horizontal axis and the H-to-C distance on the vertical axis. The z-axis of the STM imposes a mechanical constraint such that the system is confined to a diagonal line in this parameter space. The dashed grey line shows the constraint applied when the separation (the STM's z-axis) is held 100 pm further than the separation at which the barrier to reaction vanishes, and the solid grey line is the constraint when the separation is at the zero-barrier value. The green arrows represent the evolution of the system under the STM's control during approach (in the left valley) and retraction (in the deeper, energetically favored, bottom valley). The black arrow represents the energetically downhill transfer of the hydrogen from the silicon to the carbon atom.*

The system starts in the upper left of the plot, with the hydrogen atom bonded to a surface silicon atom of the H:Si(100) probe, as shown in the upper left callout. At this separation, the reaction does not proceed because the system is constrained to the dashed grey line (with $z = z^* + 100$ pm) and therefore sees an energetic barrier (the same one shown in Figure 1e). During the approach, the sample moves toward the probe (green arrows in the upper left and lower left callouts), and the barrier decreases. At a critical separation, $z^*$, shown by the solid grey line in the PES, the barrier to reaction along the line of constraint vanishes, and the reaction is uninhibited. In reality, because of quantum effects, the hydrogen atom is transferred before this point, when the barrier becomes sufficiently small. The chemical configurations before and after transfer are shown in the callout in the bottom left and bottom center. After the hydrogen is transferred, as the probe retracts (green arrows in the bottom center and bottom right callouts), the hydrogen atom remains bonded to the carbon atom. In the PES, the system remains in the deeper valley along the bottom of the plot.

The PES illustrates that when the reagents of a chemical reaction are well understood and connected to the probe and sample, the mechanical degrees of freedom can be used to steer the system through its chemical configuration space. In other words, their relative positions are controlled by the STM scanner's $x$, $y$, and $z$ degrees of freedom. While only the $z$ degree of freedom is considered in this example, all three axes could become relevant for other reactions. For more complex potential energy landscapes, the STM could be used to overcome energetic barriers or to steer the system into one of multiple wells. We expect this will enable chemical outcomes that are otherwise hard to achieve and will be useful for fundamental investigations of individual chemical reactions.

# 10. Additional Data on Hydrogen Abstraction on H:SPC

Mechanosynthetic H-abstraction is a reliable process, that can be utilized to create isolated DBs, but also to pattern DBs on the H:SPC surface. Figure S13 shows the success rates for abstracting a single hydrogen atom to create an isolated dangling bond (DB) (left) and a second hydrogen atom to form an inter-row DB (IR-DB) pair (right), using the minimum poke depth (MPD) protocol described in the main text. It should be noted that some of the interactions occurred at the first poke depth, and a shallower poke depth may have yielded the same H-abstraction result. Here, we define IR-DB as a DB pair which spans a trough separating two rows of the H:Si(100)-2x1 surface reconstruction.

Green indicates that a single hydrogen atom was successfully abstracted (27/28 times for the first DB, and 24/24 times for the second DB), where dark and light green indicate whether the DB was on-target or off-target, respectively. Purple denotes a single unsuccessful attempt, which was a large-scale de-passivation of the H-passivated SPC attributable to the EAOGe-C2 tool breaking between successive pokes, causing an "overpoke" that culminated in a large-scale depassivation event. Importantly, abstraction of two hydrogen atoms or donation of an unknown species was never observed in the cases where MPD was employed, confirming the robustness of the underlying chemistry.

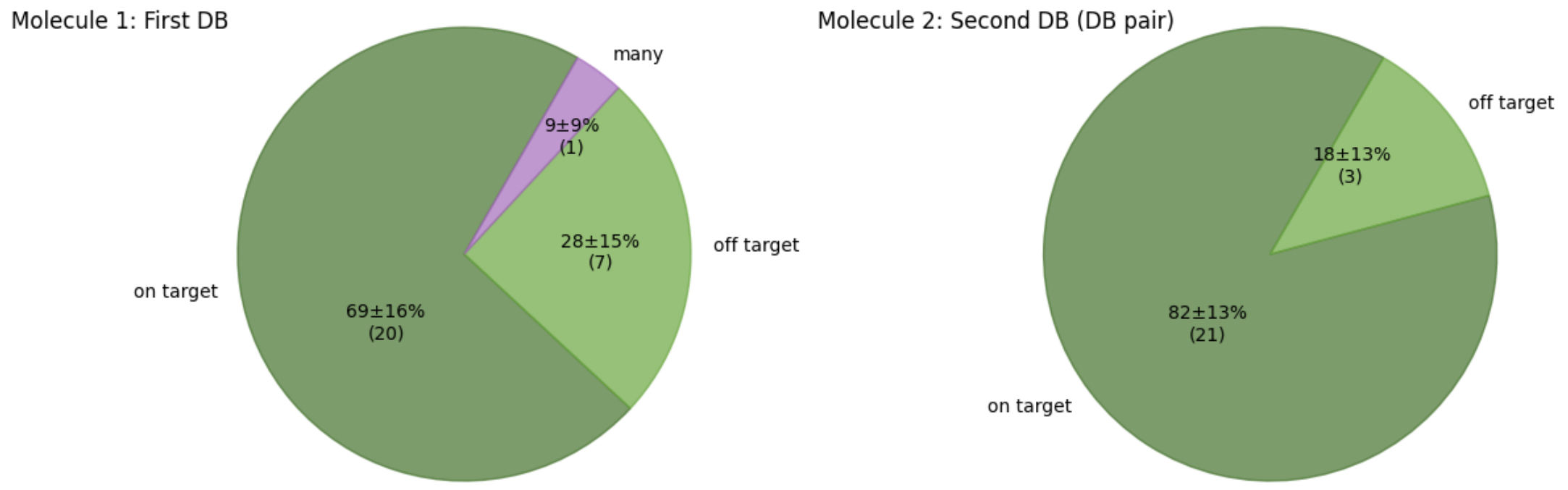

*Figure S13: Statistics gathered during early testing of DB patterning with mechanosynthetic H-abstraction to form a simple IR-DB pattern. The percentages reported are the calculated Wilson score and error with a 95% Wilson score interval used for each of the observed outcomes during the process[77] . Graph a) shows the outcomes when patterning the first DB, where 27/28 H-abstraction attempts were successful (20 were on-target, 7 were off-target). One interaction attempt was too deep, resulting in needed to re-passivate the SPC. Graph B similarly shows statistics for the second step of the IR-DB patterning, where all 24 attempts resulted in a single H-abstraction (21/24 attempts produced the desired IR-DB configuration, while 3/24 resulted in an H-abstraction that was off-target).*

Figure S14 shows two examples of IR-DBs being patterned, comprising four examples of hydrogen abstraction in total. These examples show that while the procedure is reproducible, in practice not all data sets are as ideal as that shown in Figure 4. Some of the variety of observations that we see is discussed presently.

In Figure 4, it was shown that there is a distinct variation in imaging quality between an iodine-capped molecule, an activated (deiodinated) molecule, and a hydrogen-capped molecule (EAOGe-C2I, EAOGe-C2 and EAOGe-C2H, respectively). These distinct imaging qualities provide us with feedback about the terminating species of the imager molecule and can act as a key signature of a successful mechanosynthetic reaction. However, it should be noted that even similarly capped molecular tools may present variations in imaging quality of an SPC. For hydrogen-terminated SPCs, this is particularly noticeable in the H-capped imagers, as seen in Figure S14c, g, l, and o. Further work is necessary to fully understand the variety of factors which influence these variations in imaging. Across these examples, and the one shown in Figure 4, the H-capped molecule tends to show dimer resolution, but the RPI height is reduced compared to the other molecule states considered. In Figure S14c, there is a bistability, which we sometimes observe and do not fully understand, that leads to streakiness in the image. In Figure S14k, the RPI is so faint that background noise nearly overtakes the image, making it difficult to see the RPI. Finally, in Figure S14o, an entirely different part of the SPC apex is imaged with an imaging quality consistent with an activated tool, which indicates that a nearby, possibly activated molecule is now dominating the RPI image.

DBs sometimes move to adjacent sites of the same dimer. An example is shown in Figure S14h, where the inset shows the DB in the IR-DB pattern has shifted to the adjacent positions during scanning, altering the intended IR-DB feature.

We have observed variations in the depth required to perform the mechanosynthetic hydrogen abstraction from an H:SPC. This variation may be attributed to various factors, including but not limited to: nearby features or defects on the H-terminated SPC probe, presence of nearby molecules on the sample, relative probe-sample tilt, and tunneling quality of the specific molecule, all of which will influence the absolute initial probe-sample separation at the onset of the poke.

While it is generally possible to find molecules that are stable to STM scanning with H-passivated SPCs, we have observed that some molecules can “break” during scanning or a poke sequence. A broken molecule will have an abrupt change in its imaging quality, with the resulting RPI essentially disappearing into the noise. The term “breaking” is used here very loosely. In such cases, we do not fundamentally know what change occurred, only the observed behavior.

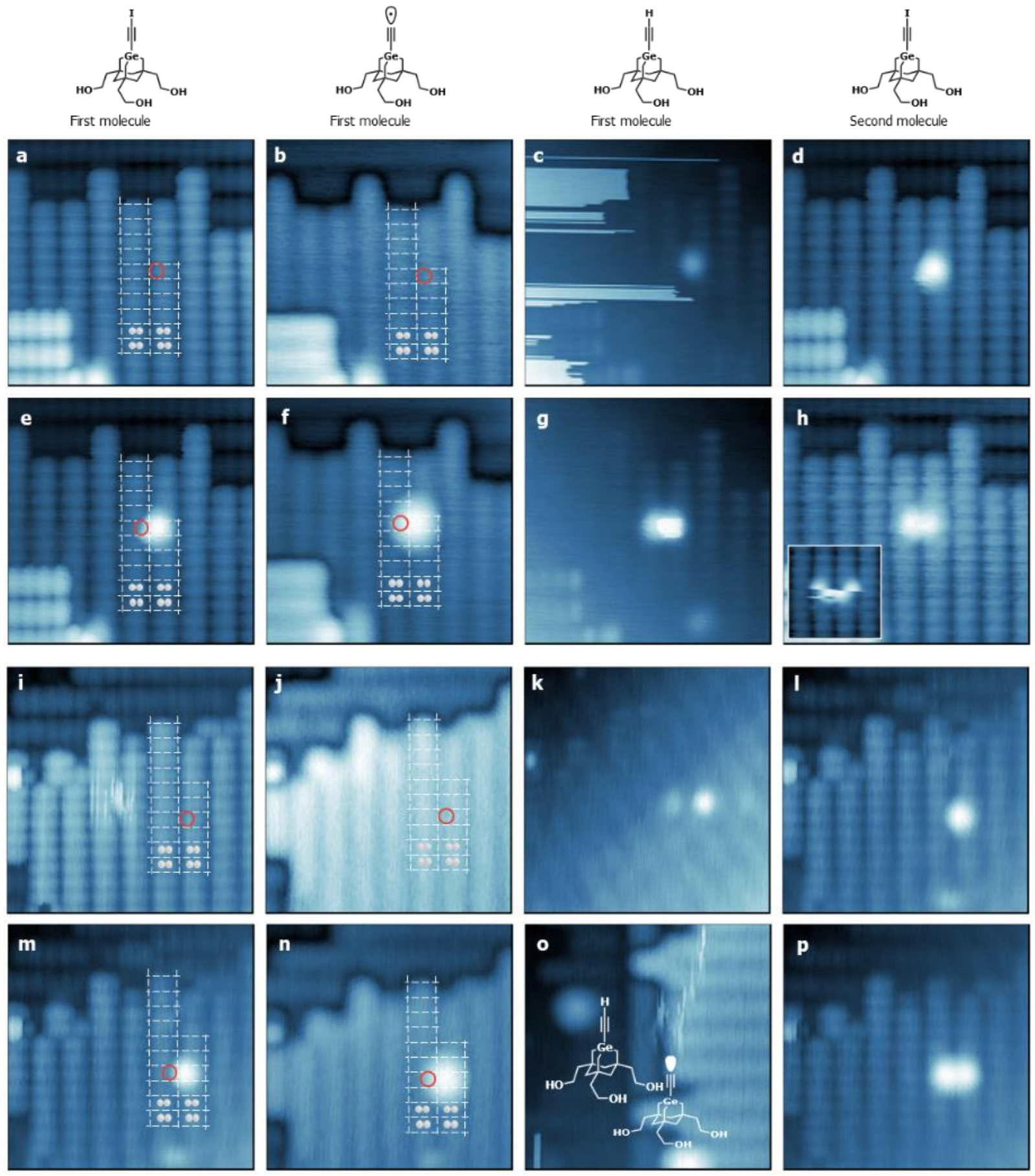

*Figure S14: Two examples of DBs being patterned via H-abstraction into inter-row configurations. Each row in the image represents a single H-abstraction step, beginning with a first iodinated EAOGe-C2I molecule (panels a, e, i and m), which is deiodinated in-situ to form EAOGe-C2 (panels b, f, j and n); a vertical poke into the desired position abstracts the hydrogen atom from the probe apex, leaving behind a presumed hydrogen capped molecule, EAOGe-C2H (panels c, g, k and o) on the sample surface and a new DB on the probe apex. Finally, a second iodine capped molecule is used to characterize the outcome (panels d, h, l, p). EAOGe-C2I images were captured at $V_B$ = 3.2V, $I_T$ = 10 pA, while EAOGe-C2 and EAOGe-C2H images were taken at $V_B$ = 3.5-3.8 V, $I_T$ = 10 pA. All scans show a 6nm x 6nm region.*

The variation in observed behavior of molecules in inverted-mode STM highlights the need for further investigation of fundamental aspects of inverted-mode STM. Nonetheless, H-abstraction and DB patterning can be performed routinely via mechanosynthesis on the probe apex, demonstrating that the technique is robust to these challenges.

Finally, while we have only shown mechanosynthetic removal of hydrogen from an H:SPC, we have also verified that it is possible to depassivate surface sites on an H:SPC via bias pulsing. Similarly to Hydrogen Depassivation Lithography (HDL) methods that have been shown in the literature, whereby a sharp metallic W or Pt tip can be used to depassivate individual sites on an H passivated sample[75,78], we can perform a similar operation in inverted mode STM using EAOGe-C2I. By positioning the EAOGe-C2I molecule over a specific site on the probe and performing a negative bias pulse (in the range of -3.4 V to -4.0 V, 20 ms duration), a hydrogen atom can be removed from the desired H:SPC site, leaving behind a DB. This method can provide an avenue for rapid patterning of an H:SPC with DBs, using a single EAOGe-C2I molecule.

One test that can confirm that the bright protrusion made after an interaction is indeed a DB is to switch it back and forth during scanning. An example of this is shown in Figure S15, using a characterizing EAOGe-C2I molecule to switch a DB. This behavior is consistent with DB switching on H:Si(100)-2x1 samples as previously reported in the literature[79–81] . This test can be done with DBs made mechanosynthetically or with bias pulses.

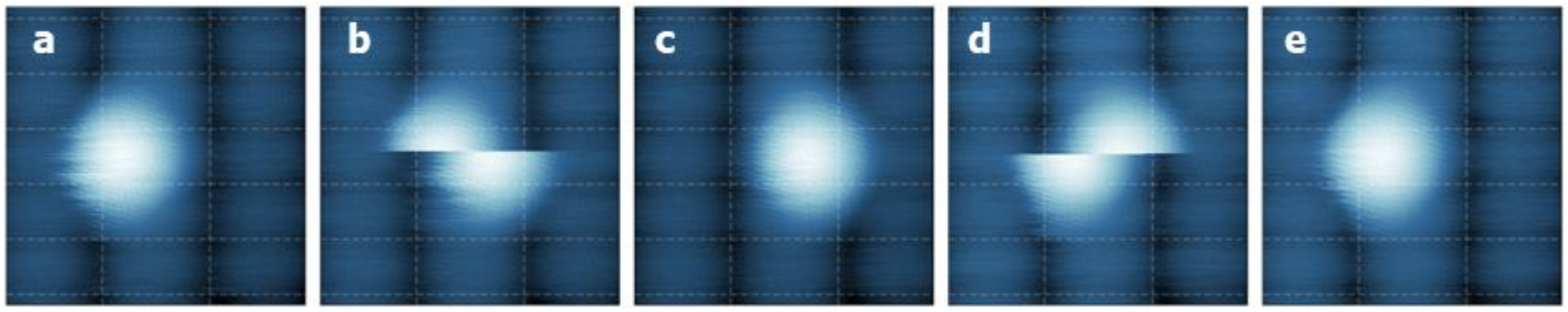

*Figure S15: Example of DB switching, which can be observed and purposefully done, via scanning with a EAOGe-C2I molecule on the surface. This serves as a confirmation that the bright protrusion on the H:SPC surface is indeed a single dangling bond, and switching can be utilized to correct off-target DBs or DB's which have moved to an adjacent surface site (see example S14, panel h). Images show a 2nm X 2nm scan area. Scan conditions are $V_S$ = 3.5V, $I_T$ = 10 pA for panels a, b and c, and $V_S$ = 3.2V, $I_T$ = 15 pA for panels d and e.*